\documentclass[11pt]{article}

\pdfoutput=1
\usepackage{graphicx}
\usepackage{amssymb}
\usepackage{amsmath}
\usepackage{bm}
\usepackage{latexsym}
\usepackage{epsfig}
\usepackage{textcomp}
\usepackage{color}
\usepackage{comment}
\usepackage{xcolor,colortbl}

\newcommand{\mc}[2]{\multicolumn{#1}{c}{#2}}
\definecolor{Gray}{gray}{0.9}
\newcolumntype{a}{>{\columncolor{Gray}}c}

\usepackage[a4paper]{geometry}
\makeatother

\begin{document}

\title{On the covariant formalism of the effective field theory of gravity and its cosmological implications}
\author{Alessandro Codello\thanks{codello@cp3-origins.net} \,and Rajeev Kumar Jain\thanks{jain@cp3.sdu.dk}\\
\emph{CP$^3$-Origins, Centre for Cosmology and Particle Physics Phenomenology}\\
\emph{University of Southern Denmark}\\
\emph{Campusvej 55, 5230 Odense M}\\
\emph{ Denmark}}
%
%
%

\date{}

\maketitle

\begin{abstract}
Following our previous work wherein the leading order effective action was computed in the covariant effective field theory of gravity, here we specialize the effective action to the FRW spacetime and obtain the effective Friedmann equations. In particular, we focus our attention on studying the cosmological implications of the non--local terms when each of them is combined with the Einstein--Hilbert action. We obtain both analytical and iterative solutions to the effective background equations in all the cases and also briefly comment on the consistency between the iterative and numerical solutions whenever possible. We find that among all the non--local terms, the imprints induced by $R\frac{1}{\square^2}R$ are very significant. Interpreting these corrections as an effective dark energy component characterized by an equation of state parameter, we find that the $R\frac{1}{\square^2}R$ correction can indeed lead to an accelerated expansion of the universe at the present epoch even in the absence of a cosmological constant. We briefly discuss some phenomenological consequences of our results. 
\end{abstract}



\section{Introduction}

Over the last few decades, cosmology has become a very precise science and it is indeed the right moment to move from phenomenology to theory, since now we finally have the possibility to falsify the proposed modifications to Einstein's theory of general relativity (GR).
While a large class of such modifications suggested recently are purely phenomenological, it is possible to predict these corrections from first principles within the framework of effective field theory (EFT), even in the absence of a complete theory of quantum gravity \cite{Donoghue:1994dn,Burgess:2003jk,Donoghue:2012zc,Donoghue:2015hwa}.
When applied to gravity, the EFT framework not only allows us to consider it together with the other forces of nature but it also manages to make distinct predictions.  When the EFT of gravity is constructed in a complete covariant approach, as we have done in our previous paper \cite{Codello:2015mba}, it is indeed possible to obtain these modifications to GR on a general background and in particular, in a cosmological setting.  

One of the major problems in cosmology is to explain the accelerated expansion of our universe at the present epoch, which is confirmed by different cosmological probes and has motivated a lot of activity in order to provide a satisfactory theoretical explanation for this observational fact \cite{Riess:1998cb,Perlmutter:1998np}. Among the plethora of possibilities, the cosmological constant $\Lambda$ is unarguably the simplest choice, however, the typical problems associated with it e.g. its enormously small value and its dominance {\it only} today has led to the exploration of other potentially viable models \cite{Weinberg:1988cp,Carroll:1991mt,Peebles:2002gy,Padmanabhan:2002ji}. It is well known that cosmic acceleration within the framework of GR can be achieved either by adding an `exotic' matter component to GR or by the modification of GR itself. As a result, these alternative scenarios can be broadly classified into two main categories: i) scalar field models \cite{Wetterich:1987fm,Ratra:1987rm} and ii)  modified gravity models \cite{Sotiriou:2008rp,DeFelice:2010aj,Clifton:2011jh}. In the former case, either a minimally coupled scalar field with a suitably flat potential is added to Einstein's gravity or it is more entwined with gravity as, for instance, in the scalar-tensor theories \cite{Brans:1961sx,Damour:1992we}. In the latter, the modifications to gravity could be local as in $f(R)$ models as well as non--local type. Both of these broad possibilities have been studied to a great extent in the literature, for some recent reviews, see  \cite{Copeland:2006wr,Jain:2010ka}. 

As outlined earlier, the EFT of gravity leads to consistent (local and non--local) modifications of GR which could turn out to very relevant in different contexts and some of them can also  explain dark energy. The distinction we want to make is that EFT induced corrections are exactly determined by computation, leaving no space for further tuning of the parameters.  By adapting to a Friedman--Robertson--Walker (FRW) background and using the leading order (LO) effective action computed in \cite{Codello:2015mba}, we start the investigation of the cosmological implications of the covariant EFT of gravity. In particular, we analyze the influence of the leading low energy non--local corrections to the background expansion and understand their role in the evolution of the universe at late times.

The imprints of non--local terms have been considered in cosmology in many different contexts, for instance, as candidates of dark energy to explain the present acceleration of the universe \cite{Espriu:2005qn,Deser:2007jk,Jhingan:2008ym,Koivisto:2008xfa,Barvinsky:2011hd,Barvinsky:2011rk,Solodukhin:2012ar,Maggiore:2013mea,Ferreira:2013tqn,Woodard:2014iga,Maggiore:2014sia,Maggiore:2015rma}, as a mechanism for screening of the cosmological constant \cite{Nojiri:2010pw,Bamba:2012ky}, to obtain accelerating solutions by means of a time delay in the Friedmann equations \cite{Choudhury:2011xf}, for a resolution of the cosmological singularity \cite{Biswas:2010zk,Li:2015bqa}, in the Newtonian limit \cite{Koivisto:2008dh,Conroy:2014eja} and also to obtain the spherically symmetric solutions \cite{Bronnikov:2009az,Kehagias:2014sda,Dirian:2014xoa}. Accelerating solutions have also been obtained in a general class of non--local higher derivative gravity \cite{Capozziello:2008gu,Elizalde:2011su,Myrzakulov:2014hca}.
While this indicates that the non--local terms indeed lead to very rich and interesting phenomenology, they have mostly been studied as modifications to GR, and also argued as being consistent in some cases but have not been obtained in an effective covariant framework. 

This paper is organized as follows: in the following section, we quickly recap the EFT of gravity, write down the effective action to the quadratic order in curvature in the Weyl basis and specialize it to the FRW background. In section \ref{localg}, we study the classical theory as well as the leading local correction in the form of $R^2$ and obtain the analytical and numerical solutions. In section \ref{nlocalg}, we study the leading non--local terms that arise at the second order in curvature and obtain the analytical, iterative and numerical solutions for all the cases. In section \ref{comp}, we compare the these corrections in terms of the equation of state parameter for dark energy.  Finally, in section \ref{discuss}, we summarize our results and conclude with some discussions and a future outlook. In the appendix, we discuss some subtleties associated with the Green's function in the FRW spacetime and outline how to obtain the kernel for the logarithmic distribution.


\section{Effective action from the covariant EFT of gravity}\label{eftg}

In the previous paper \cite{Codello:2015mba}, we had computed the leading order effective action to the second order in the curvature expansion within the covariant EFT of gravity. The final result for the gravitational part of the effective action is given by
\begin{eqnarray} \label{ea_R2_Final}
\!\!\!\! \Gamma & = & \frac{1}{16\pi G}\int d^{4}x\sqrt{-g}\left(R-2\Lambda\right)-\frac{1}{\xi}\int d^{4}x\sqrt{-g}\, R^{2}-\frac{1}{2\lambda}\int d^{4}x\sqrt{-g}\, C^{2}\nonumber \\
 &  & -\int d^{4}x\sqrt{-g}\, R\, \mathcal{F}\!\left(\frac{-\square}{m^2}\right)\!R-\int d^{4}x\sqrt{-g}\, C_{\mu\nu\alpha\beta}\, \mathcal{G}\! \left(\frac{-\square}{m^2}\right)\!C^{\mu\nu\alpha\beta}+O(\mathcal{R}^{3})\,,
\end{eqnarray}
where $G, \Lambda, \xi$ and $\lambda$ are phenomenological parameters which must be fixed by the experiments or observations while the structure functions $\mathcal{F}$ and $\mathcal{G}$ are completely determined once the matter content of the theory is specified. The general form of these structure functions can be found in \cite{Codello:2015mba}.
In order to understand the implications of this effective action in a cosmological setting, we work in a $(3+1)$--dimensional, spatially flat, FRW universe described by the line element
\begin{equation}
ds^{2}=-dt^{2}+a^{2}(t)d{\bf x}^{2}\,,\label{CT_1.1}
\end{equation}
where $t$ is the cosmic time and $a(t)$ is the scale factor.
For this metric  the Weyl tensor vanishes identically and therefore, the Weyl part of the effective action (\ref{ea_R2_Final}) will not contribute to the background equations of motion (EOM). 
However, this is not true when dealing with cosmological perturbations and the contributions due to these terms must be taken into account which could lead to very distinct signatures in the cosmological observables.
As discussed in \cite{Codello:2015mba}, at third order in the curvature expansion, the contributions from the conformal anomaly will also be present which can have interesting cosmological implications (see for instance \cite{Shapiro:2008sf,Mottola:2010gp} and references therein) but we will not discuss them here.
\renewcommand{\arraystretch}{2}
\begin{table}[t]
\begin{centering}
\begin{tabular}{cccca}
\hline\hline \\[-1.15cm]
\mc{1}{}  & \mc{1}{0} & \mc{1}{$\frac{1}{2}$} & \mc{1}{1} & \mc{1}{2} \tabularnewline
\hline 
$m^{2}$ & $m_{\phi}^{2} = V''(v)$ & $m_{\psi}^{2}$ & $\mu^{2}$ & $-2\Lambda+16 \pi G\,V(v)$  \tabularnewline
\hline 
$\alpha$ & $\frac{(1-\chi)^2}{144}$ & 0 & 0 & $\frac{1}{8}$\tabularnewline
\hline 
$\beta$ & $\frac{14-18 \chi+3 \chi ^2}{216}$ & $\frac{2}{27}$ & $-\frac{13}{108}$ & $\frac{4}{27}$ \tabularnewline
\hline 
$\gamma$ & $-\frac{1-\chi ^2}{72}$ & $-\frac{1}{36}$ & $\frac{1}{18}$ &  $\frac{49}{36}$ \tabularnewline
\hline 
$\delta$ & $-\frac{15-6 \chi-\chi ^2}{144}$ & $\frac{1}{6}$ & 0 & $\frac{17}{24}$ \tabularnewline\hline\hline
\end{tabular}
\par\end{centering}
\caption{Coefficients of the non--local terms for various spins taken from \cite{Codello:2015mba}. Spin $0$ fields are conformally coupled for $\chi=1$ and minimally coupled for $\chi=0$. Spins $\frac{1}{2}$ and $1$ are minimally coupled while the last column for spin $2$, shaded in grey, are indeed the predictions of the EFT of gravity. All the numbers in the table must be divided by $(4 \pi)^2$. In case, one considers $N$ particles of a given matter species, the numbers in the first three columns should be multiplied by $N_{0}, N_{1/2}$ and $N_{1}$, respectively while the last column remains unchanged. 
\label{NLHK}}
\end{table}

In this paper, as we are only interested in studying the background evolution, we can safely neglect the Weyl terms in (\ref{ea_R2_Final}) and the total effective action including the matter action $S_{\textrm{m}}$ then simplifies to
\begin{equation}
\Gamma=\frac{1}{16\pi G}\int d^{4}x\sqrt{-g}\left(R-2\Lambda\right)-\frac{1}{\xi}\int d^{4}x\sqrt{-g}\, R^{2}-\int d^{4}x\sqrt{-g}\, R\,\mathcal{F}\!\left(\frac{-\square}{m^2}\right)\!R+S_{\textrm{m}}\,,\label{EOM_1}
\end{equation}
where the quadratic $R^2$ term is a local term while the structure function $\mathcal{F}$ is non--local in the low energy massless limit, i.e. $m^2 \ll -\square$, and has the following form
\begin{equation}
\mathcal{F}\!\left(\frac{-\square}{m^2}\right)  =  \alpha\log\frac{-\square}{m^{2}} +\beta\frac{m^{2}}{-\square} +\gamma\frac{m^{2}}{-\square}\log\frac{-\square}{m^{2}}+\delta\left(\frac{m^{2}}{-\square}\right)^2+...\label{EOM_2}
\end{equation}
The coefficients $\alpha,\beta,\gamma$ and $\delta$ are indeed {\it predictions} of the EFT of gravity which ultimately depend only on the field content of the theory and are listed in Table \ref{NLHK}. In the case of scalars and fermions, $m$ is the effective mass of the particle. In the case of photons only the massless limit can be considered $m^2 \rightarrow \mu^2$, i.e. only the first logarithmic term survives, while in the case of gravity, the effective mass is  $m^2=-2\Lambda+16\pi G\, V(v)$. In the absence of a cosmological constant and/or an effective potential, we add an IR regularization mass $\mu^2$ as in the case of massless fields.

In a cosmological context, the first logarithmic term, the only one present in the massless limit, has been studied in \cite{Espriu:2005qn,Cabrer:2007xm} and in the framework of EFT in \cite{Donoghue:2014yha}. The successive $\frac{1}{-\square}$ term is part of the class of non--local modifications of gravity first proposed in \cite{Deser:2007jk} while the last term in (\ref{EOM_2}) has been proposed and studied in \cite{Maggiore:2013mea,Maggiore:2014sia}. Apart from the leading logarithmic term, the other non--local terms were considered at a phenomenological level argued as consistent non--local modifications of gravity, but as shown in \cite{Codello:2015mba}, all these terms arise in the framework of the  EFT of gravity and all of them must be taken into account together to construct a viable scenario of the universe. As mentioned earlier, the coefficients of these terms are not free parameters and one must tune the matter content of the theory for a given term to be the leading one. Below, we discuss the conditions under which this happens. 
\begin{itemize}
\item
For the case of massless matter (without the graviton), only the scalar leading logarithmic term is present as shown in the Table \ref{NLHK}. However, it vanishes in the conformal case $\chi=1$. Therefore, there is no correction from the structure function $\mathcal{F}$ and the leading term in the action will be the one induced by the conformal anomaly.   
\item
The case of massive matter (without the graviton) with $\chi=1$ again eliminates the logarithmic term thereby making the leading contribution being $\frac{1}{-\square}$. Also, if one takes $N_{1/2}=2 N_{1}$, all the logarithmic contributions vanish i.e. $\alpha=\gamma=0$.
\item
One can find a condition for a theory (without the graviton) such that a particular coefficient vanishes. For instance, for $\chi=0$  and $12 N_{0}=3 N_{1/2}=4 N_{1}$, one gets $\beta=0$ while for $N_{0}=2 N_{1/2}=2 N_{1}$, one gets $\gamma=0$. However, it is not possible to have both $\beta$ and $\gamma$ vanish simultaneously.  Also, for $\chi=1$, one can obtain $\beta=0$ for $15 N_{0}=6 N_{1/2}=10 N_{1}$ and $\gamma=0$ for $N_{1/2}=2 N_{1}$. Moreover, for $\chi=1$, it is instead possible to kill both $\beta$ and $\gamma$ by choosing $N_{0}=3 N_{1/2}=6 N_{1}$ such that the leading correction is given by $\frac{1}{\square^2}$.
\item
In the presence of the graviton, it is not possible to make the leading logarithmic corrections go away, i.e. $\alpha=0$, as the contributions due to both the scalars and gravitons are positive. 
\item
Including the graviton,  it is possible to get $\beta=0$ by choosing $165 N_{0}=11 N_{1/2}=15 N_{1}$ for $\chi=0$  and by choosing $88 N_{0}=11 N_{1/2}=16 N_{1}$ for $\chi=1$. One can also attain $\gamma=0$ by choosing $25 N_{0}= N_{1/2}=50 N_{1}$ for $\chi=0$ and $2 N_{1/2}=53 N_{1}$ for $\chi=1$. Furthermore, it is possible to make both $\beta$ and $\gamma$ vanish simultaneously only for $\chi=1$.
\end{itemize}

Along the same logic presented above, one can also find a condition such that the correction term with $\delta$ vanishes with or without the graviton. It is also evident from this analysis that when the graviton contribution is taken into account, it is not possible to make either $\frac{1}{-\square}$ or $\frac{1}{\square^2}$ the leading correction since the logarithmic term will always be present.

In this paper, even if we have just shown that all the non--local terms should be considered together, we start by studying the solutions corresponding to different cases where each of these non--local terms is combined with the Einstein--Hilbert term separately and later, compare them with each other. 
For each case, we write down the covariant form of the field equations and obtain their solutions. In general, when a correction term is present in a theory along with the classical counterpart, the modified Einstein's equation can be written as
\begin{equation}\label{modee}
G_{\mu\nu}+\Delta G_{\mu\nu} = 8 \pi G\, T_{\mu\nu}\,,
\end{equation}
where $\Delta G_{\mu\nu}$ is the correction term arising from the local or the non--local terms. Note that, the correction term $\Delta G_{\mu\nu}$ is covariantly conserved.


\section{Local theory}\label{localg}

In this section, we first review the background solutions in the classical theory described by the Einstein--Hilbert action  (with the cosmological constant $\Lambda$) and matter. Later we study the solutions for $R^2$ gravity which is the leading local correction to the classical theory  within the EFT of gravity.


\subsection{Classical theory}

At the classical level, the theory is described by the Einstein--Hilbert action (with the cosmological constant $\Lambda$)
\begin{equation}
\Gamma=\frac{1}{16\pi G}\int d^{4}x\sqrt{-g}\left(R-2\Lambda\right)+S_{\textrm{m}}\,.\label{CT_1}
\end{equation}
The Friedmann EOM can be obtained by varying this action
\begin{eqnarray}
H^{2}&=&\frac{\Lambda}{3}+\frac{8\pi G}{3} \rho  \label{CT_5} \\
\frac{\ddot a}{a} &=&  \frac{\Lambda}{3} - \frac{4\pi G}{3} (\rho+3p) \,,\label{CT_6}
\end{eqnarray}
where $H \equiv {\dot a}/a$ is the Hubble parameter and $\rho$ and $p$ indicate the energy density and pressure of the matter content, respectively. Conservation of the energy--momentum tensor leads to the continuity equation
\begin{equation}
\dot \rho + 3 H (1+w) \rho = 0 \,,\label{CT_7}
\end{equation}
where $w=p/\rho$ is the equation of state parameter of the matter content. Note that only two among the equations (\ref{CT_5}), (\ref{CT_6}) and (\ref{CT_7}) are independent. In a universe dominated by the cosmological constant with $w=-1$, the solutions to these equations lead to a de Sitter expansion, for which the scale factor is given by
\begin{equation}\label{dsa}
a_{\Lambda}(t)  = e^{\sqrt{\frac{\Lambda}{3}}\, (t-t_0) }\,.
\end{equation}
For the case of radiation ($w=1/3$) with $\rho_{r} \propto a^{-4}$, we get 
\begin{equation}\label{ra}
a_{r}(t) = \left(\frac{t}{t_0}\right)^{1/2}\,,
\end{equation}
while for the case of matter or dust ($w=0$)  with $\rho_{m} \propto a^{-3}$, one finds 
\begin{equation}\label{ma}
a_{m}(t) = \left(\frac{t}{t_0}\right)^{2/3}\,,
\end{equation}
where the normalization is fixed by imposing $a(t_0)=1$. In the later sections we will evaluate the leading order corrections on these background solutions. 

\subsection{$R^{2}$ gravity}

As explained earlier, the local quadratic $R^2$ term appears at the leading order in the EFT of gravity (\ref{EOM_1}). This is one of the most interesting and well studied example of a local higher order correction to Einstein's theory which was originally studied in \cite{Starobinsky:1980te}. The action in this case is
\begin{equation}
\Gamma=\frac{1}{16\pi G}\int d^{4}x\sqrt{-g}\left(R-2\Lambda\right)-\frac{1}{\xi}\int d^{4}x\sqrt{-g}\, R^{2}+S_{\textrm{m}}\,,\label{ST_1}
\end{equation}
where $\xi$ is a free parameter which must be fixed by, for instance, the condition that inflation is driven by this term.  However, one can also think of the usual situation where inflation is induced by the matter sector and the $R^2$ term is not present at all. 
Although it is possible to study the dynamics of the action (\ref{ST_1}) in the Einstein frame by doing a conformal transformation, we rather choose to work in the Jordan frame as it will be useful for the case when this local term is combined with the non--local terms to study the dynamics of the full effective action (\ref{EOM_1}).  

To obtain the correction term to the Einstein's equations, we make a variation of the second term in the action (\ref{ST_1}) using the following relations
\begin{equation}
\delta\sqrt{-g}  =  \frac{1}{2}\sqrt{-g}\, h \qquad\qquad
\delta R  =  -h_{\mu\nu}R^{\mu\nu}+\nabla^{\mu}\nabla^{\nu}h_{\mu\nu}-\square h \,,\label{V_1}
\end{equation}
where $h_{\mu\nu}= \delta g_{\mu \nu}$. The resulting expression becomes
\begin{eqnarray}\label{eoms_1}
-\frac{\xi}{16\pi G}\Delta G_{\mu\nu} & = & 2\left(R_{\mu\nu}-\frac{1}{4}g_{\mu\nu}R\right)R-2 \left(\nabla_{\mu}\nabla_{\nu}-g_{\mu\nu}\square\right) R \,.
\end{eqnarray}
By inserting the FRW metric in this equation and using the relations
\begin{equation}
R_{00} = -3 (\dot H+H^2)  \qquad\qquad R=6(\dot H + 2H^2)\,,
\end{equation}
in (\ref{modee}) leads to the modified Friedmann EOM for $R^{2}$ gravity
\begin{equation}
H^{2}-\frac{96\pi G}{\xi}\left(2H\ddot{H}+6H^{2}\dot{H}-\dot{H}^{2}\right)=\frac{\Lambda}{3}+\frac{8\pi G}{3}\rho \,,\label{ST_6}
\end{equation}
which can be immediately compared with \cite{Starobinsky:1980te, Capozziello:2014hia}. Note that, the inclusion of the correction term makes this differential equation of third order for the scale factor instead of first order in its absence.


In the absence of matter, the EOM for $R^{2}$ gravity is given by
\begin{equation}
\alpha H^{2}+2H\ddot{H}+6H^{2}\dot{H}-\dot{H}^{2}=0 \qquad\qquad\alpha=-\frac{\xi}{96\pi G}\,.\label{ST_7}
\end{equation}
This is a second order non linear differential equation which is not exactly solvable. In the following, we will first find exact analytical solutions for the pure $R^{2}$ gravity and later obtain approximate and numerical solutions to this equation. It is useful to note that the classical theory is recovered  as $\alpha \rightarrow \infty$. 

In the case of pure $R^{2}$ gravity $(\alpha=0)$ we can find an analytical solution which for particular choices of the integration constants reduces to both de Sitter and radiation. We want to solve
\begin{equation}
2H\ddot{H}+6H^{2}\dot{H}-\dot{H}^{2}=0\,.\label{ST_8}
\end{equation}
Clearly $H=H(t_0)$ is the constant solution to this equation which corresponds to the de Sitter space thus pure $R^{2}$ gravity inflates, as is well known. To find the other solutions, we first shift to a new variable $Z=\dot{H}/\sqrt{H}$ to find
\begin{equation}
dZ=-3\sqrt{H}dH\,,
\end{equation}
which can be easily integrated to give the following implicit solution
\begin{equation}
t+C_{2}=\int\frac{dH}{C_{1}\sqrt{H}-2H^{2}}\,,\label{ST_9}
\end{equation}
depending on the two integration constants $C_{1}$ and $C_{2}$. The integral in (\ref{ST_9}) can be performed exactly but the resulting expression is not very illuminating apart from the case $C_{1}=C_{2}=0$, yielding 
\begin{equation}
H=\frac{1}{2\,t}\,,
\end{equation}
which is indeed the radiation solution (\ref{ra}). The fact that radiation is an exact solution of pure $R^{2}$ gravity is expected since $R=0$ for radiation in FRW and any action with more than one power of $R$ will lead to EOM that admits such a solution. The de Sitter solution is recovered in the limit $C_{2}\to-\infty$ when $C_{1}\neq0$. 
\begin{figure}[!t]
\begin{center}
\includegraphics[width=7.8cm,height=5.8cm]{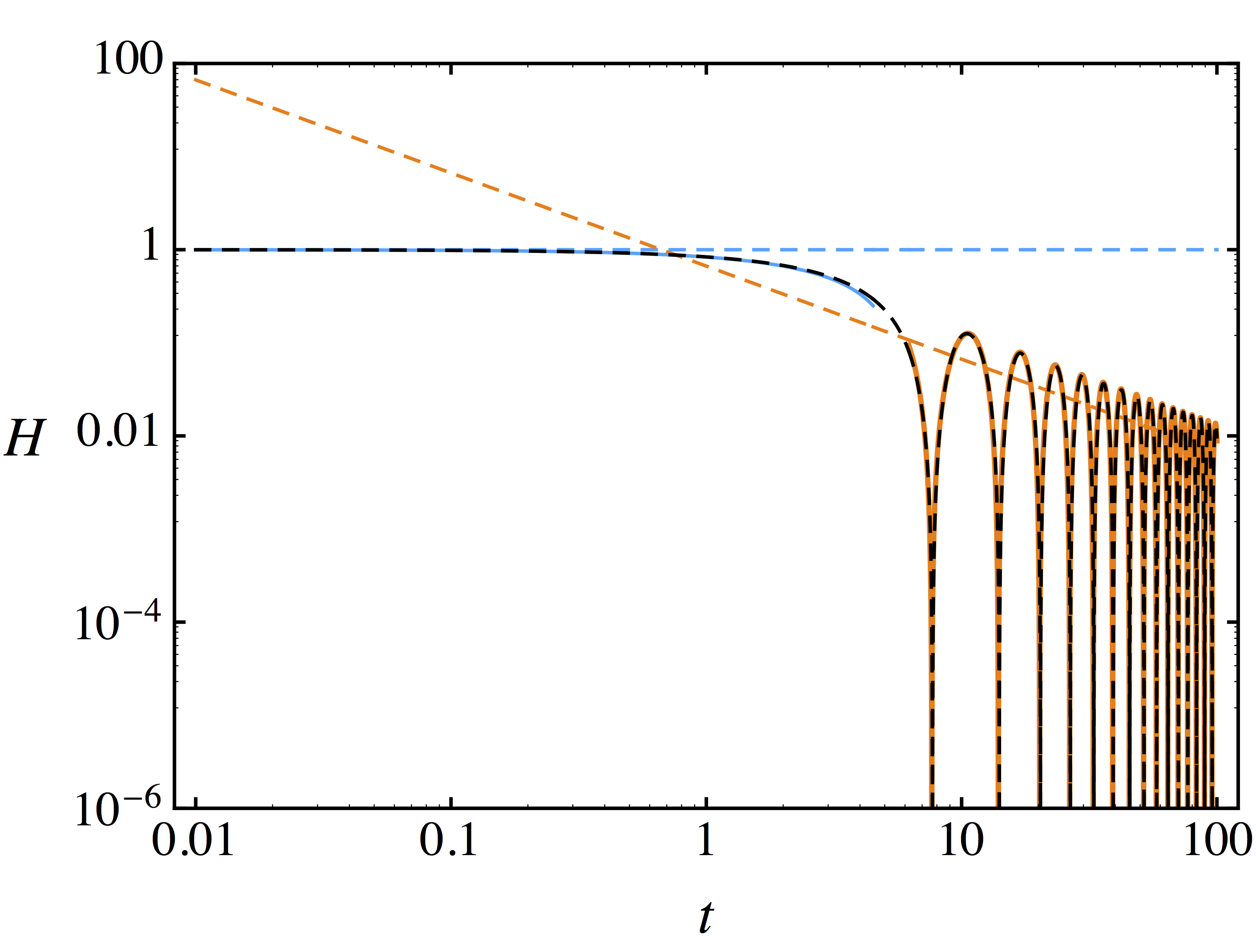}
\hskip 2pt
\includegraphics[width=7.8cm,height=5.8cm]{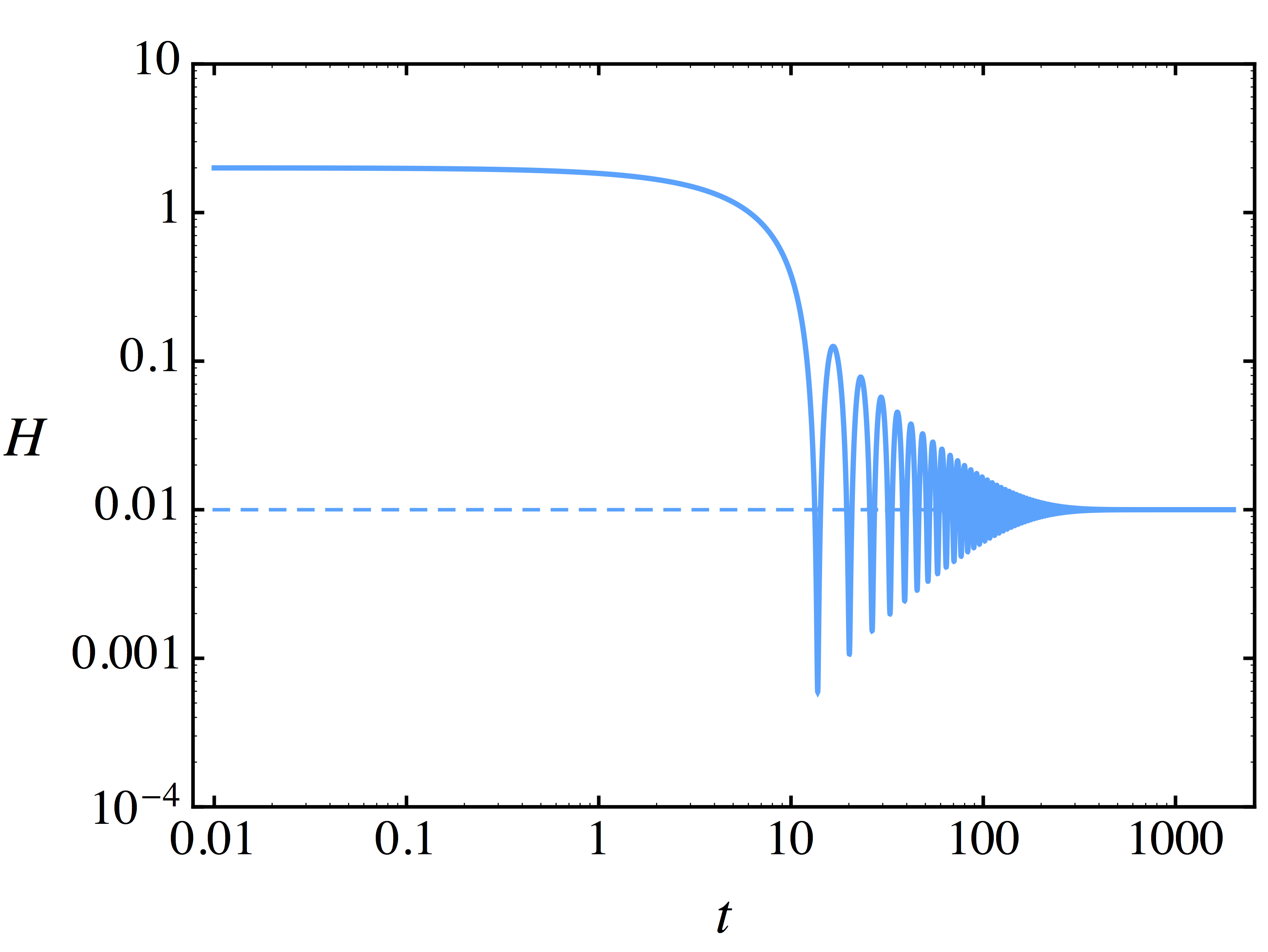}
\end{center}
\caption{On the left, the time evolution of the Hubble parameter $H$ is plotted for $R^{2}$ gravity without matter. The dashed black line is the full numerical solution to (\ref{ST_7}). The dashed blue line indicates the constant solution for $H$ normalized to $H(t_0)$ while the solid blue line indicates the quasi de Sitter solution described in the text. The solid orange line represents the damped oscillating solution at late times and the dashed orange line indicates the slope $\frac{2}{3t}$ as in the case of matter.  On the right, the time evolution of $H$ for $R^{2}$ gravity including $\Lambda$ is plotted as  shown by the solid blue line. The presence of $\Lambda$, as indicated by the dashed blue line, suppresses the damped oscillating solution, leading to a late time de Sitter phase.}
\label{ST_H1}
\end{figure}

As mentioned earlier, we can not solve exactly $R^{2}$ gravity in the absence of matter but we can try to understand the solutions in different regimes when certain terms in (\ref{ST_7}) can be neglected as compared to the others. If at early times $H$ is nearly constant, we can ignore the higher order time derivative terms $2H\ddot{H}$ and $\dot{H}^{2}$ as compared to $6H^{2}\dot{H}$ in (\ref{ST_7}), one finds the quasi de Sitter regime with the solution given by $H(t) \simeq H(t_0)-\frac{\alpha}{6}(t-t_0)$. On the contrary, when $H$ is time dependent, the cubic term $6H^{2}\dot{H}$ can be dropped, leading to an oscillatory solution \cite{Arbuzova:2011fu}. As shown in Figure \ref{ST_H1}, these oscillations are indeed present when equation (\ref{ST_7}) is integrated numerically.
In the presence of matter, it is easy to realize that $R^2$ gravity with cosmological constant or radiation is exactly solved by de Sitter and radiation solutions, respectively.


We now turn our attention to solve $R^{2}$ gravity numerically. We first solve it without any matter content and then solve it in the presence of $\Lambda$ as well as with radiation and matter. For these cases, we solve the EOM (\ref{ST_6}) together with the continuity equation (\ref{CT_7}). For $R^2$ gravity without matter, we find that the exact numerical solution shows a quasi de Sitter phase followed by a decaying oscillatory behavior and is well corroborated by the approximate solutions in different regimes discussed earlier, as shown in Figure {\ref{ST_H1}. The effect of adding the cosmological constant is to damp the oscillations at late times before reentering the de Sitter phase. In Figure {\ref{ST_H2}, we show the solutions for $R^{2}$ gravity with radiation as well as with matter and radiation. In both cases, the oscillatory phase damps out and the solution enters into a radiation phase. Note that the global behavior of these solutions is qualitatively very interesting from a phenomenological point of view.

\begin{figure}[!t]
\begin{center}
\includegraphics[width=7.8cm,height=5.8cm]{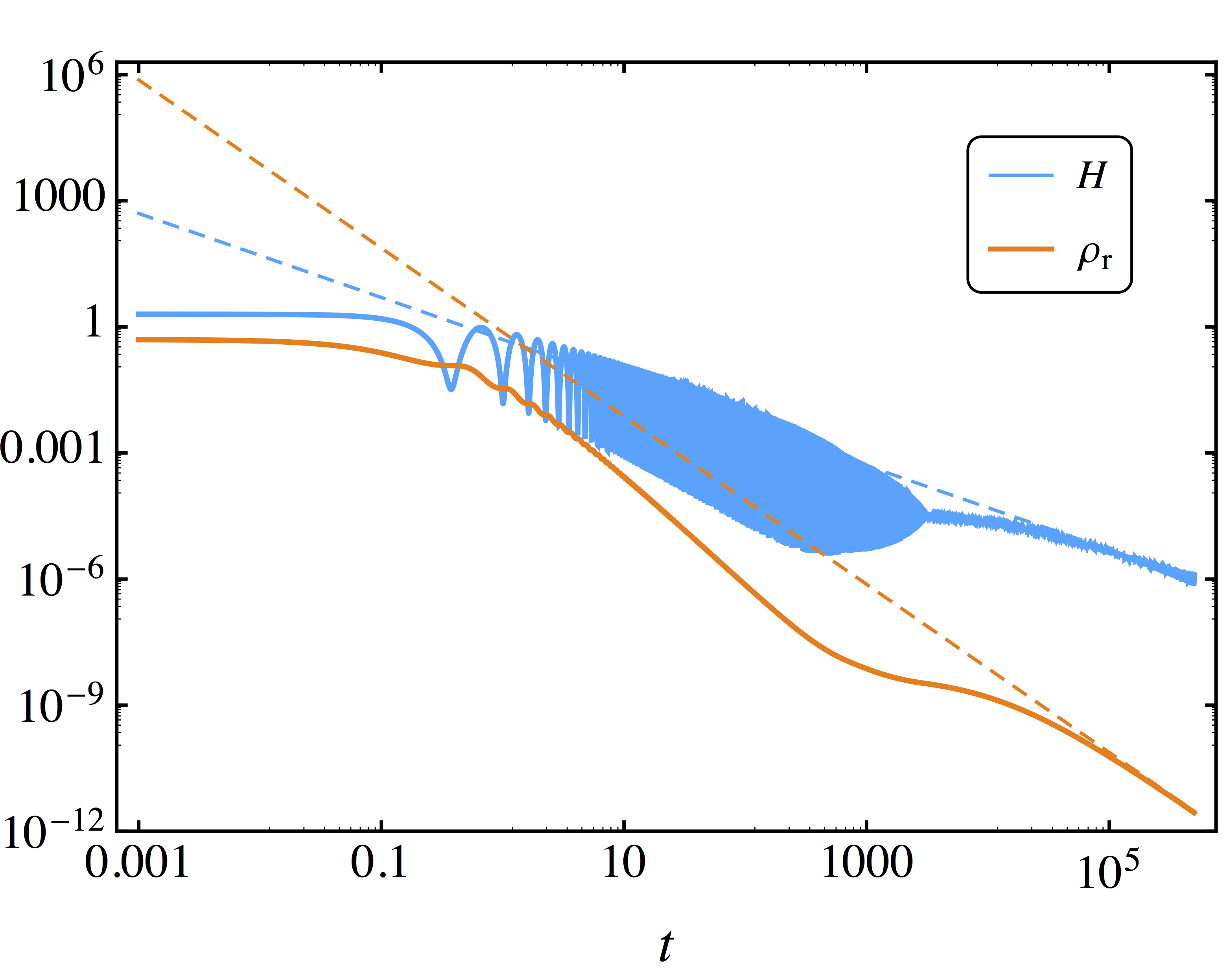}
\hskip 2pt
\includegraphics[width=7.8cm,height=5.8cm]{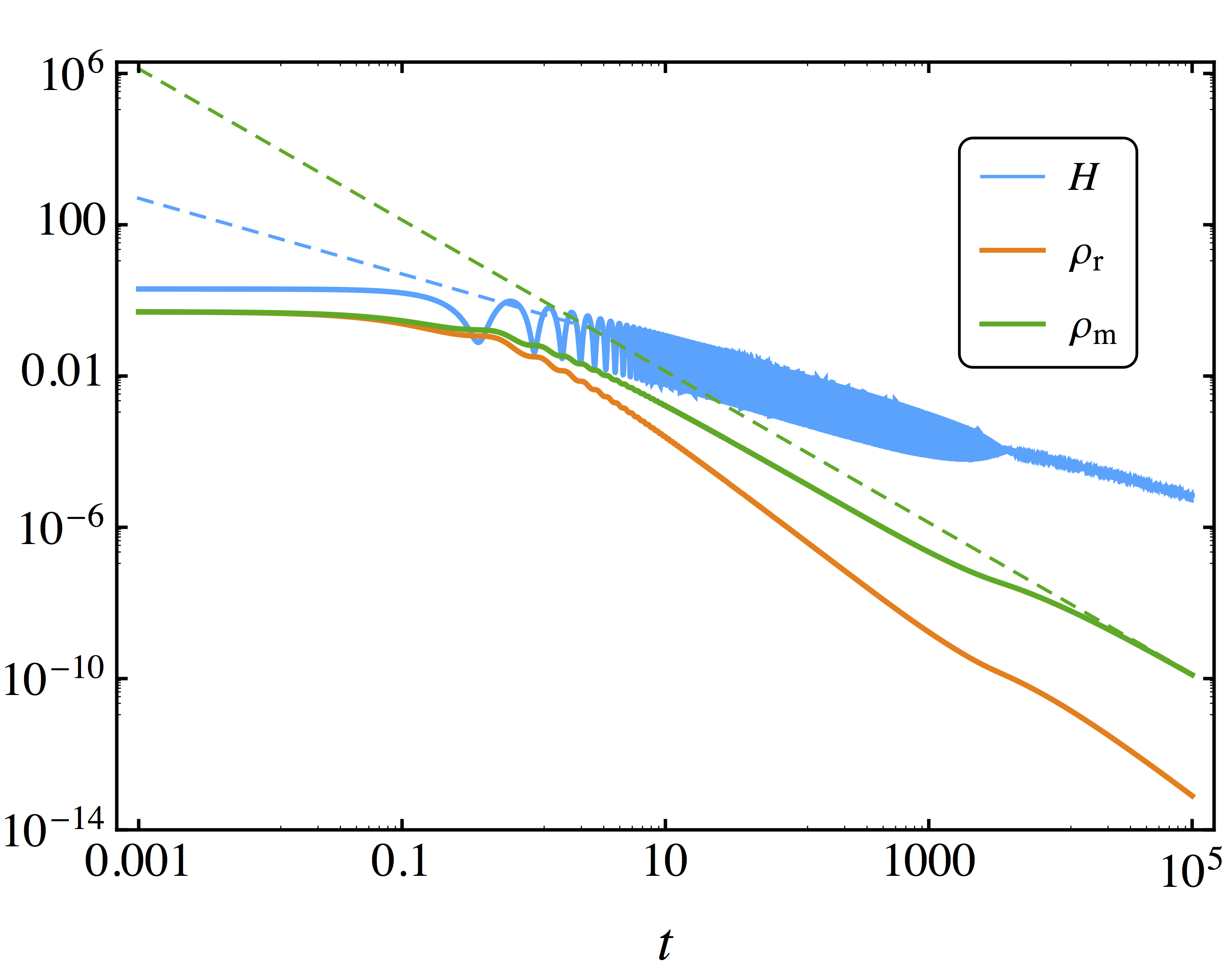}
\end{center}
\caption{On the left, the time evolution of the Hubble parameter $H$ is plotted for $R^{2}$ gravity together with radiation. After the oscillatory solution damps out, both $H$ and the radiation energy density $\rho_r$ asymptote to $\frac{1}{2t}$ and $\frac{3}{4t^2}$, as shown by the dashed blue and dashed orange lines, respectively. In the right panel, the time evolution of $H$ is plotted for $R^{2}$ gravity  with both radiation and matter present. Also for this case, as expected, after the end of the oscillatory phase, $H$, $\rho_r$ and the matter energy density $\rho_m$ nicely settle to their asymptotic behaviors, as indicated by the various dashed lines, respectively. }
\label{ST_H2}
\end{figure}


\section{Non--local theory}\label{nlocalg}

In this section, we shall combine the Einstein--Hilbert action with different non--local terms present in (\ref{EOM_2}), study the solutions of the relative effective Friedmann EOM and discuss their relevance for the late time evolution of the universe.


\subsection{{\small $R\log\frac{-\square}{\mu^{2}}R$} gravity}\label{donoghue}

As discussed in our previous paper \cite{Codello:2015mba}, the leading non--local correction appearing at the second order in curvatures in the EFT of gravity is given by the logarithmic term which has been studied in \cite{Espriu:2005qn,Cabrer:2007xm} and in the framework of EFT in \cite{Donoghue:2014yha}.
In this section, we will study the effects of the non--local action of the form
\begin{equation}
I_{\rm NL}= \int d^{4}x\sqrt{-g}\, R\log\frac{-\square}{\mu^{2}}R\,.\label{D_2}
\end{equation}
As discussed in section \ref{eftg}, this correction term is only generated by minimally coupled scalars and gravitons while it vanishes for fermions, photons and conformally coupled scalars. Before proceeding, we first clarify the precise meaning of the logarithmic term in (\ref{D_2}) which is given by the following integral representation
\begin{equation}
\log\frac{-\square}{\mu^{2}} \equiv \int_{0}^{\infty}ds\left[\frac{1}{\mu^{2}+s}-\frac{1}{-\square+s}\right]\,.\label{D_2.1}
\end{equation}
This definition allows us to express the logarithm in terms of the Green's function of the operator $-\square +s$, from which it will inherit the boundary conditions.
We now provide a covariant derivation of the EOM for this case. Upon variation, the action in (\ref{D_2}) leads to 
\begin{equation}
\delta I_{\rm NL}=\int d^{4}x \left\{ \delta\sqrt{-g}\,R\log\frac{-\square}{\mu^{2}}R + 2\sqrt{-g}\,\delta R\log\frac{-\square}{\mu^{2}}R+ \sqrt{-g}\,R\, \delta\log\frac{-\square}{\mu^{2}} R\right\} \,.
\end{equation}
The first two terms can be easily dealt with using the variations given in (\ref{V_1}) while for the last term, we vary (\ref{D_2.1}) to arrive at 
\begin{equation}
\delta\log\frac{-\square}{\mu^{2}} = \int_{0}^{\infty}\! ds\,\frac{1}{-\square+s}\delta\square\frac{1}{-\square+s}\,,
\end{equation}
where the variation of $\square$ acting on a scalar is given by 
\begin{equation}
\delta\square\phi = -h_{\mu\nu}\nabla^{\mu}\nabla^{\nu}\phi-\nabla^{\mu}h_{\mu\nu}\nabla^{\nu}\phi+\frac{1}{2}\nabla^{\mu}h\nabla_{\mu}\phi\,.
\label{var_box}
\end{equation}
Using these relations, it is now easy to extract the modification to the Einstein's tensor
\begin{eqnarray*}
-\frac{1}{16\pi G\alpha}\Delta G_{\mu\nu} & = & 2\left(R_{\mu\nu}-\frac{1}{4}g_{\mu\nu}R\right)\!\!\left(\log\frac{-\square}{\mu^{2}}R\right)-2 \left(\nabla_{\mu}\nabla_{\nu}-g_{\mu\nu}\square\right) \left(\log\frac{-\square}{\mu^{2}}R\right) \\
 & -& \!\int_{0}^{\infty}\! ds\left[-\nabla_{\mu}\left(\frac{1}{-\square+s}R\right)\nabla_{\nu}\left(\frac{1}{-\square+s}R\right)\right.\\
 & + & \! \left. \frac{1}{2}g_{\mu\nu}\left\{\nabla^{\alpha}\!\left(\frac{1}{-\square+s}R\right)\!\nabla_{\alpha}\!\left(\frac{1}{-\square+s}R\right)\!
 +\left(\frac{1}{-\square+s}R\right)\!\square\!\left(\frac{1}{-\square+s}R\right)\right\}\right]\!.
\end{eqnarray*}
When dealing with non--local corrections of this type, it turns out to be very useful to introduce auxiliary fields to localize them at the expense of increasing the number of equations. In this particular case, we define $L=\log\frac{-\square}{\mu^{2}}R$ and $U_{s}=\frac{1}{-\square+s}R$ to obtain
\begin{eqnarray}\label{eomd_1}
-\frac{1}{16\pi G\alpha}\Delta G_{\mu\nu} & = & 2\left(R_{\mu\nu}-\frac{1}{4}g_{\mu\nu}R\right)L-2 \left(\nabla_{\mu}\nabla_{\nu}-g_{\mu\nu}\square\right) L \nonumber \\
 &  & -\int_{0}^{\infty}\! ds\left[-\nabla_{\mu}U_{s}\nabla_{\nu}U_{s}+\frac{1}{2}g_{\mu\nu}\nabla^{\alpha}U_{s}\nabla_{\alpha}U_{s}+\frac{1}{2}g_{\mu\nu}U_{s}\square U_{s}\right]\\ 
L & = & \int_{0}^{\infty}ds\left[\frac{1}{\mu^{2}+s}R-U_{s}\right]\\
(-\square+s)U_{s} & = & R\,.
\end{eqnarray}
Note that, while the first line of (\ref{eomd_1}) is equivalent to (\ref{eoms_1}) after the substitution $L\to R$, the integral term in the second line appears due to the variation of the logarithm. 

We are interested in obtaining the correction term (\ref{eomd_1}) in the FRW background. The time--time component of the correction term is
\begin{eqnarray}
\!\!\!\!-\frac{1}{16\pi G\alpha}\Delta G_{00} & = & -\left(\frac{1}{2} R-6H^2\right) L +6 H \dot L -\frac{1}{2} \int_{0}^{\infty}\! ds\left[-{\dot U_{s}}^2+U_{s} \ddot U_{s}+3 H U_{s} {\dot U_{s}}\right],
\label{eomd_2}
\end{eqnarray}
where the first two terms are evaluated using the following expression for the field $L$
\begin{equation}
L(t) = \int dt' a^3(t') L_{-}^{\rm FRW}(t-t') R(t')\, ,
\label{Lfrw}
\end{equation}
and the kernel $L_{-}^{\rm FRW}$ is the retarded representation of the operator $\log\frac{-\square}{\mu^{2}}$ in the FRW background. Ignoring the time dependence of the scale factor it can be written as
\begin{equation}
L_{-}^{\rm FRW}(t-t')  =  \frac{L^{\rm flat}_{-}(t-t')}{\sqrt{a^{3}(t)a^{3}(t')}}+O(\dot{a})\,, \label{lfrw}
\end{equation}
such that the FRW kernel is just the flat space kernel 
\begin{equation}
L^{\rm flat}_{-}(t-t')  =  -2\lim_{\epsilon\rightarrow0}\left[\frac{\theta(t-t'-\epsilon)}{t-t'}+\delta(t-t')\log\mu\,\epsilon\right]\,,
\label{D_3}
\end{equation}
except the time dependent normalization. The details of this derivation can be found in appendix \ref{appA}.
For the moment, we also neglect the last integral term in equation (\ref{eomd_2}), which appears due to the variation of the logarithmic term. Note that, this term is a higher order term (in time derivatives) as compared to the first two terms. The resulting EOM by including this term will be a non-trivial integro-differential equation. For our purpose, to access these quantum corrections and to keep the calculation tractable as much as possible, we have used the approximation wherein the higher order time derivative terms can be ignored. We now  evaluate the first two terms by inserting (\ref{Lfrw}) together with (\ref{lfrw}) and (\ref{D_3}) into (\ref{eomd_2}) which leads to the following effective Friedmann EOM
\begin{eqnarray}\label{D_6}
H^{2}(t)&+&16\pi G\alpha \left(\frac{1}{6} R(t)+H^2(t)\right) a^{-\frac{3}{2}}(t)\int dt'\, a^{\frac{3}{2}}(t')R(t')\, L^{\rm flat}_{-}(t-t') \nonumber \\
&-&32 \pi G \alpha H(t)\, a^{-\frac{3}{2}}(t) \int dt'\, L^{\rm flat}_{-}(t-t')\,\frac{d}{dt'}\left(a^{\frac{3}{2}}(t')R(t')\right)=\frac{8\pi G}{3}\rho(t)\,,
\end{eqnarray}
which can be immediately compared with \cite{Donoghue:2014yha}. Although the correction term appears non--local in time due to the retarded boundary conditions imposed, it is unambiguous and perfectly causal. However, the effective EOM (\ref{D_6}) is an integro--differential equation and as such it is generally difficult to solve.

Radiation is an exact but trivial solution since for $R=0$ the correction term in (\ref{D_6}) vanishes (as in the case of $R^{2}$ gravity). This is only true in the Weyl basis that we are working in and will not be noticed in a non--Weyl basis. 
Less trivially, the correction term also vanishes for the de Sitter expansion (\ref{dsa}), since now the two integrals in (\ref{D_6}) compensate each other. Thus de Sitter is also an exact solution of the effective Friedmann EOM (\ref{D_6}), although, again, this will not be true in a different (i.e. non--Weyl) basis for the non--local terms. 

Matter or dust instead is not an exact solution as the correction terms in (\ref{D_6}) do not vanish for this case, and so we treat it iteratively. Since we are only interested in the LO corrections in this paper, we can treat them as perturbations above the unperturbed background and use the classical solution as an input to calculate them in an iterative manner. Following this, if we insert the matter Friedmann solution (\ref{ma}) in (\ref{D_6}), we find
\begin{equation}
H^{2}(t)-\frac{256\pi G\alpha}{3\,t^{4}}\left[\log\mu t+\log\left(\frac{t}{t_{0}}-1\right)+\frac{2}{3}\left(\frac{t}{t_{0}}-1\right)\right]=\frac{8\pi G}{3}\rho_{m}(t_{0})\left(\frac{t_{0}}{t}\right)^{2}\,.\label{D_8}
\end{equation}
This iterative procedure converts the integro--differential equation into a simple differential equation \cite{Donoghue:2014yha}. We have assumed $t>t_{0}$, with $t_{0}$ being a reference time, such that $a(t_{0})=1$ to solve this equation.
Notice that, due to the weaker time dependence of the correction term, it remains subdominant with respect to the classical solution and therefore, will not affect the asymptotic evolution of the universe. The coefficient $\alpha$, from Table \ref{NLHK}, is given by
\begin{equation}
\alpha=\frac{1}{(4 \pi)^2} \left[\frac{(1-\chi)^2}{144}N_0+\frac{1}{8}\right]\,,\label{D_9}
\end{equation}
where $N_0$ is the number of scalars and $\chi=0\,(1)$ for minimally (conformally) coupled fields. Note that for a reasonable number of scalars the graviton contribution to $\alpha$ is dominant. To obtain the numerical solutions of (\ref{D_8}), we have set $8\pi G=1$ and $\mu=1$. In Figure {\ref{D_a-H-1}}, we have plotted the time evolution of the scale factor and the Hubble parameter for the classical solution and the LO corrections for the case of $N_0=100$ minimally coupled scalars. As is evident from the figure, the magnitude of the LO corrections remains exceedingly small.  With even larger number of scalars, one can expect that the magnitude of the LO corrections will increase but the character of the effect will not change. Moreover, as the scale factor evolves, the LO corrections die away due to their subdominant time dependence and the evolution is given by the usual classical solution. 
Although we have not attempted a complete numerical solution for this case, the analysis presented in the later sections indicate that the iterative solution indeed captures the subdominant effects. 

\begin{figure}[!t]
\includegraphics[width=7.8cm,height=5.6cm]{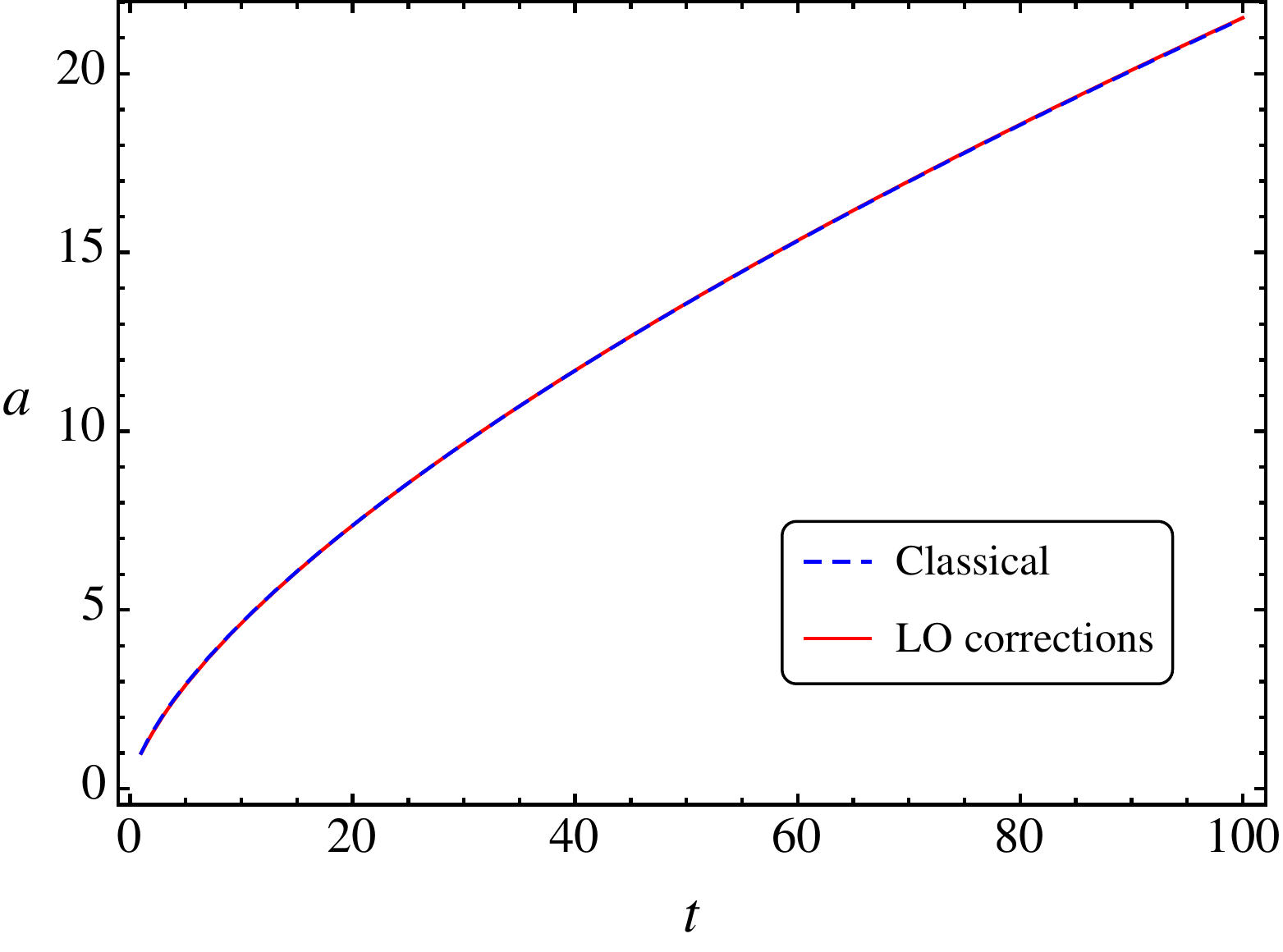}
\hskip 8pt
\includegraphics[width=7.8cm,height=5.9cm]{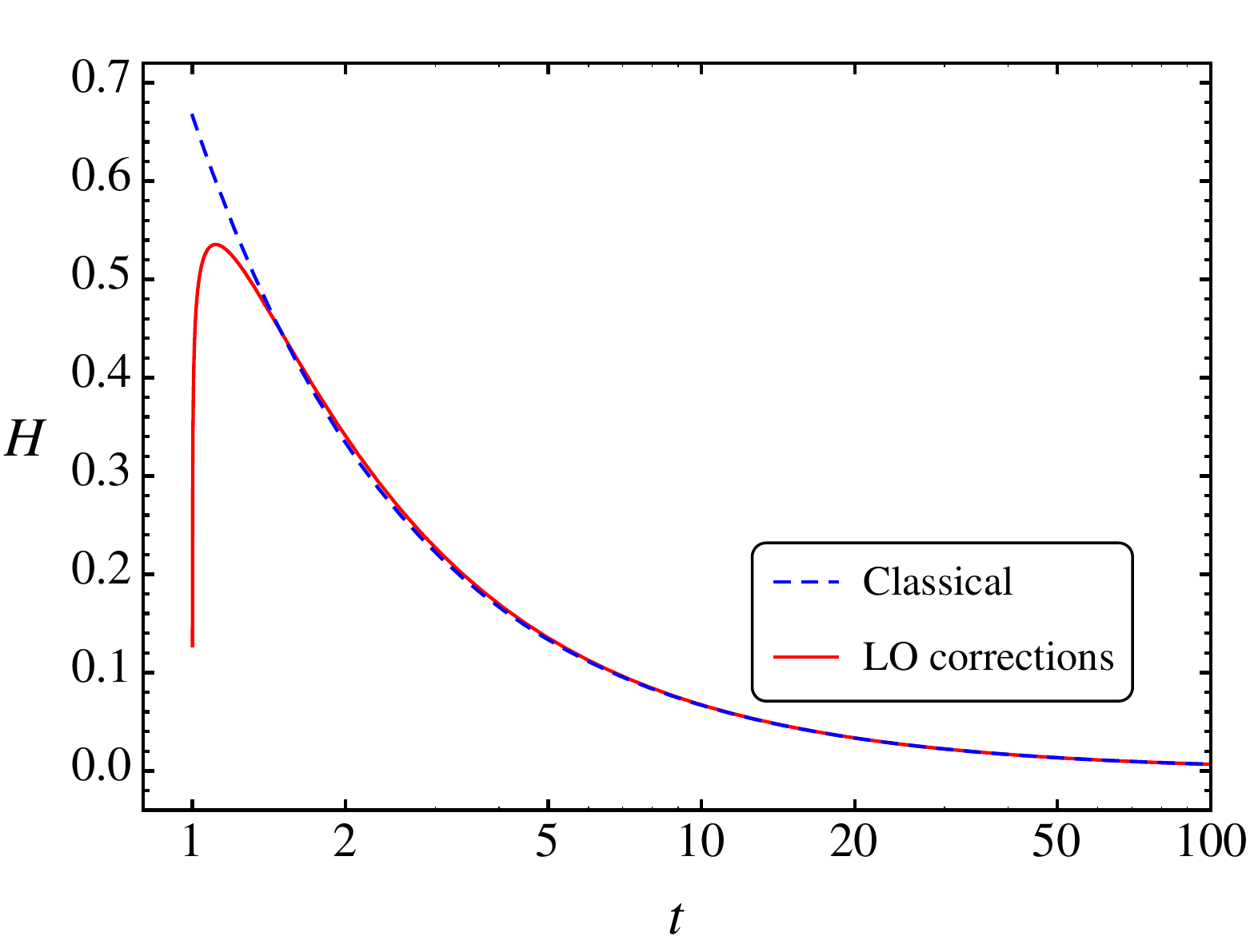}
\caption{The time evolution of the scale factor (left) and the Hubble parameter (right) for the case of the logarithmic correction term in an expanding matter dominated universe with 100 minimally coupled scalars ($\chi=0$) and the graviton.}
\label{D_a-H-1}
\end{figure}



\subsection{{\small $R\frac{1}{-\square}R$} gravity} \label{woodard}

This correction term has been studied in the cosmological background, for the first time, in \cite{Deser:2007jk}. In this case, the non--local action is 
\begin{equation}
I_{\rm NL}=-\int d^{4}x\sqrt{-g}\, R\frac{1}{\square}R\,.
\label{wood}
\end{equation}
It is evident that this correction term represents a long range interaction between the scalar curvatures and is generally induced by all types of matter content, viz. scalars, fermions, photons and gravitons as clearly seen from Table \ref{NLHK}.
To derive the covariant EOM, we start with a variation of (\ref{wood}) with respect to the metric
\begin{equation}
\delta I_{\rm NL}=-\int d^{4}x\left\{ \delta\sqrt{-g}\,R\frac{1}{\square}R+2\sqrt{-g}\,\delta R\frac{1}{\square}R-\sqrt{-g}\,R\frac{1}{\square}\delta\square\left(\frac{1}{\square}R\right)\right\} \,,
\end{equation}
where we have used the exact relation $\delta\frac{1}{\square}=-\frac{1}{\square}\delta\square\frac{1}{\square}$ formally obtained by varying the identity $\square\frac{1}{\square}=1$.
Using the variations (\ref{V_1}) and (\ref{var_box}) we immediately find the following correction to the Einstein's tensor
\begin{eqnarray} 
\frac{1}{16\pi G\beta m^{2}}\Delta G_{\mu\nu} & = & 2G_{\mu\nu}\left(\frac{1}{\square}R\right)+2g_{\mu\nu}R-2\nabla_{\mu}\nabla_{\nu}\left(\frac{1}{\square}R\right)\nonumber\\
 &  & +\nabla_{\mu}\left(\frac{1}{\square}R\right)\nabla_{\nu}\left(\frac{1}{\square}R\right)-\frac{1}{2}g_{\mu\nu}\nabla^{\alpha}\left(\frac{1}{\square}R\right)\nabla_{\alpha}\left(\frac{1}{\square}R\right)\,.
 \label{eomw_1}
\end{eqnarray}
As a check, note that this relation, if evaluated in $d=2$ (i.e. when $G_{\mu\nu}=0$), gives exactly the $\left\langle T_{\mu\nu}\right\rangle $ of Polyakov's action \cite{Codello:2010mj}. It also agrees with the correction term discussed in \cite{Deser:2007jk}.
As done in the previous case, we can linearize the correction terms in (\ref{eomw_1}) by introducing an auxiliary field,  $U=\frac{1}{-\square}R$, to find
\begin{eqnarray*}
\frac{1}{16\pi G\beta m^{2}}\Delta G_{\mu\nu} & = & -2G_{\mu\nu}U+2g_{\mu\nu}R+2\nabla_{\mu}\nabla_{\nu}U+\nabla_{\mu}U\nabla_{\nu}U-\frac{1}{2}g_{\mu\nu}\nabla^{\alpha}U\nabla_{\alpha}U\\
-\square U & = & R\,.
\end{eqnarray*}
It is now straightforward to write down the effective Friedmann EOM
\begin{eqnarray}
H^{2}+16\pi G\beta m^{2}\left(\frac{1}{6}\dot U^{2}-2H \dot U-2H^{2}U\right) & = & \frac{8\pi G}{3}\rho \label{eomw_2_1}\\
\ddot U+3H\dot U & = & 6\left(2H^2+\dot H\right)\,.
\label{eomw_2_2}
\end{eqnarray}
Contrary to the case of the logarithmic correction term, the EOM in this case are a simple system of ordinary differential equations. Together with the continuity equation (\ref{CT_7}), they form a closed system which can be solved for a given matter content of the theory.


Radiation with $R=0$ is an exact solution of these EOM while de Sitter and matter are not. In these latter cases, we shall solve the equations (\ref{eomw_2_1}) and (\ref{eomw_2_2}) iteratively, as done earlier. For the de Sitter expansion given in (\ref{dsa}), we first solve (\ref{eomw_2_2}) to get the solution for $U$ with the boundary conditions $U(t_0)={\dot U}(t_0)=0$, and insert it in (\ref{eomw_2_1}) to obtain the correction term. The effective Friedmann EOM is then given by  
\begin{equation}
H^2 - \frac{128 \pi G}{9}  \beta m^2 \Lambda \left(1 +\sqrt{3 \Lambda}\,(t-t_0)-  e^{-2 \sqrt{3 \Lambda }\, (t-t_0)}\right) = \frac{\Lambda}{3}\,.\label{W_1}
\end{equation}
This iterative procedure reduces two equations to one effective equation which can now be easily solved for given initial conditions. The second term in the above equation arises due to the non--local correction which contains a constant term, a linearly growing term as well as an exponentially decaying term in time. Note that, the correction term vanishes as $t \to t_0$ as well as $\Lambda \to 0$, as expected while at late times, i.e. $t \gg t_{0}$, the exponentially decaying term  can be safely neglected compared to the other two terms. It is also clear that the characteristic time scale involved is $1/\sqrt{3\Lambda}$. In order to solve this equation, we first need to read off the expression for $\beta$ from Table \ref{NLHK} which is
\begin{equation}
\beta=\frac{1}{(4 \pi)^2} \left[\frac{14-18\chi+3 \chi^2}{216}N_0+\frac{2}{27}N_{1/2}-\frac{13}{108}N_{1}+\frac{4}{27}\right]\,.\label{W_2}
\end{equation}
To obtain the numerical solution of (\ref{W_1}) we have set $8\pi G=1$, $m^2=1$ and $\Lambda=1$ and we focus on scalars and graviton contributions to $\beta$, although it can be easily done including fermions and photons. 
\begin{figure}[!t]
\includegraphics[width=7.6cm,height=5.5cm]{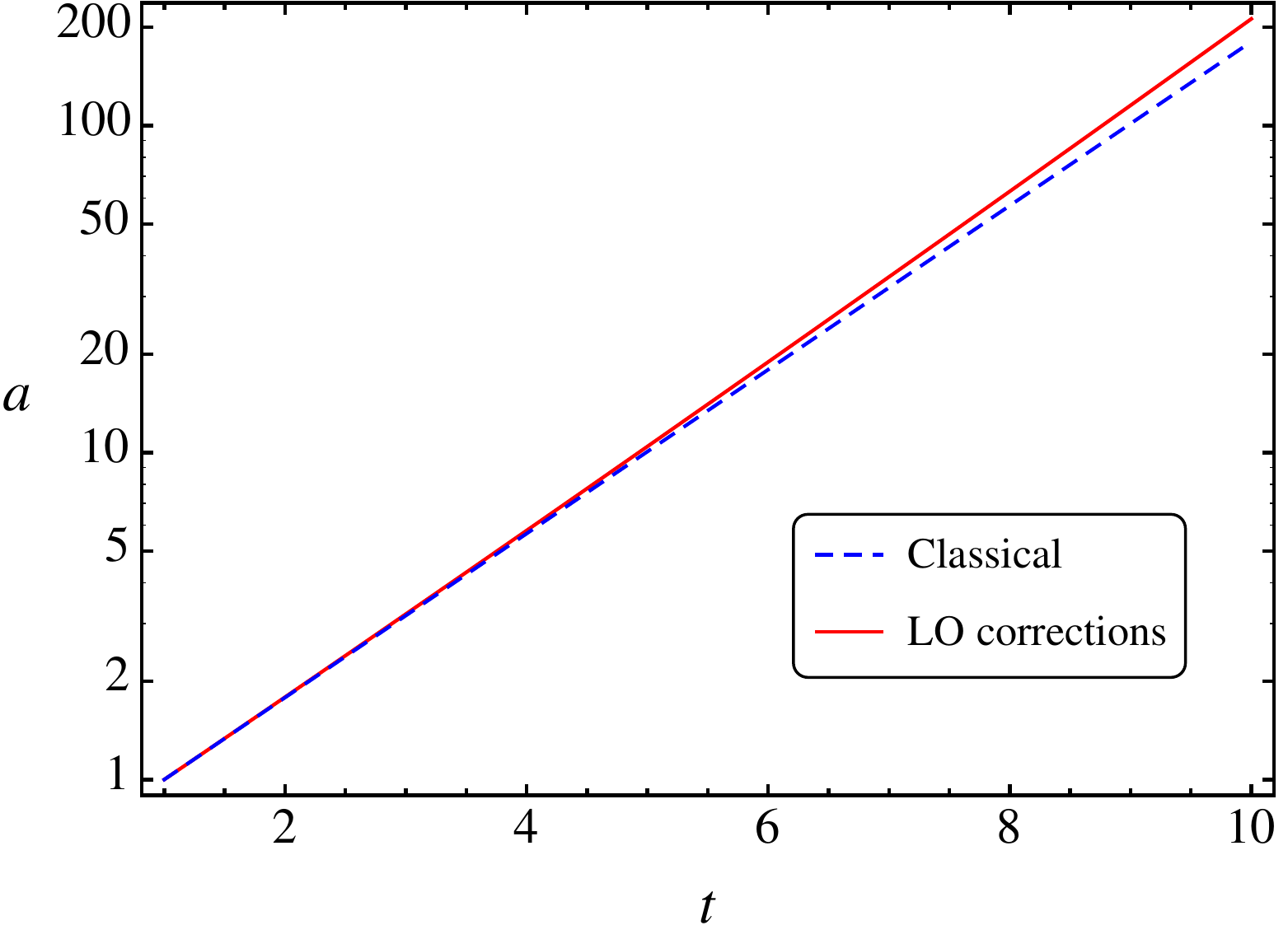}
\hskip 8pt
\includegraphics[width=7.6cm,height=5.5cm]{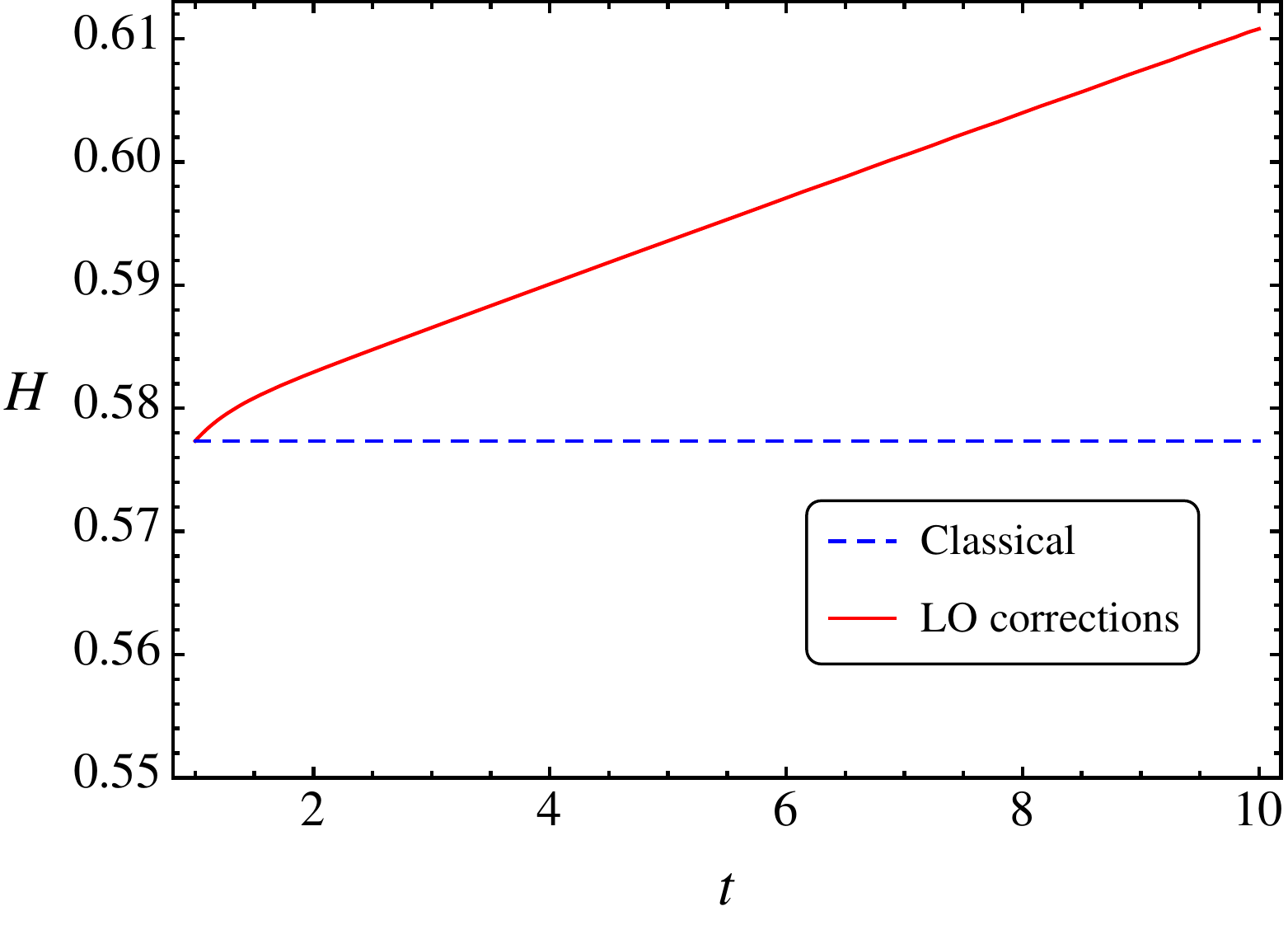}
\caption{The time evolution of the scale factor (left) and the Hubble parameter (right) for the case of $R\frac{1}{-\square}R$ correction term in a de Sitter universe with a single minimally coupled scalar ($\chi=0$) and the graviton.}
\label{W_a-H-1}
\vskip 8pt
\includegraphics[width=7.6cm,height=5.5cm]{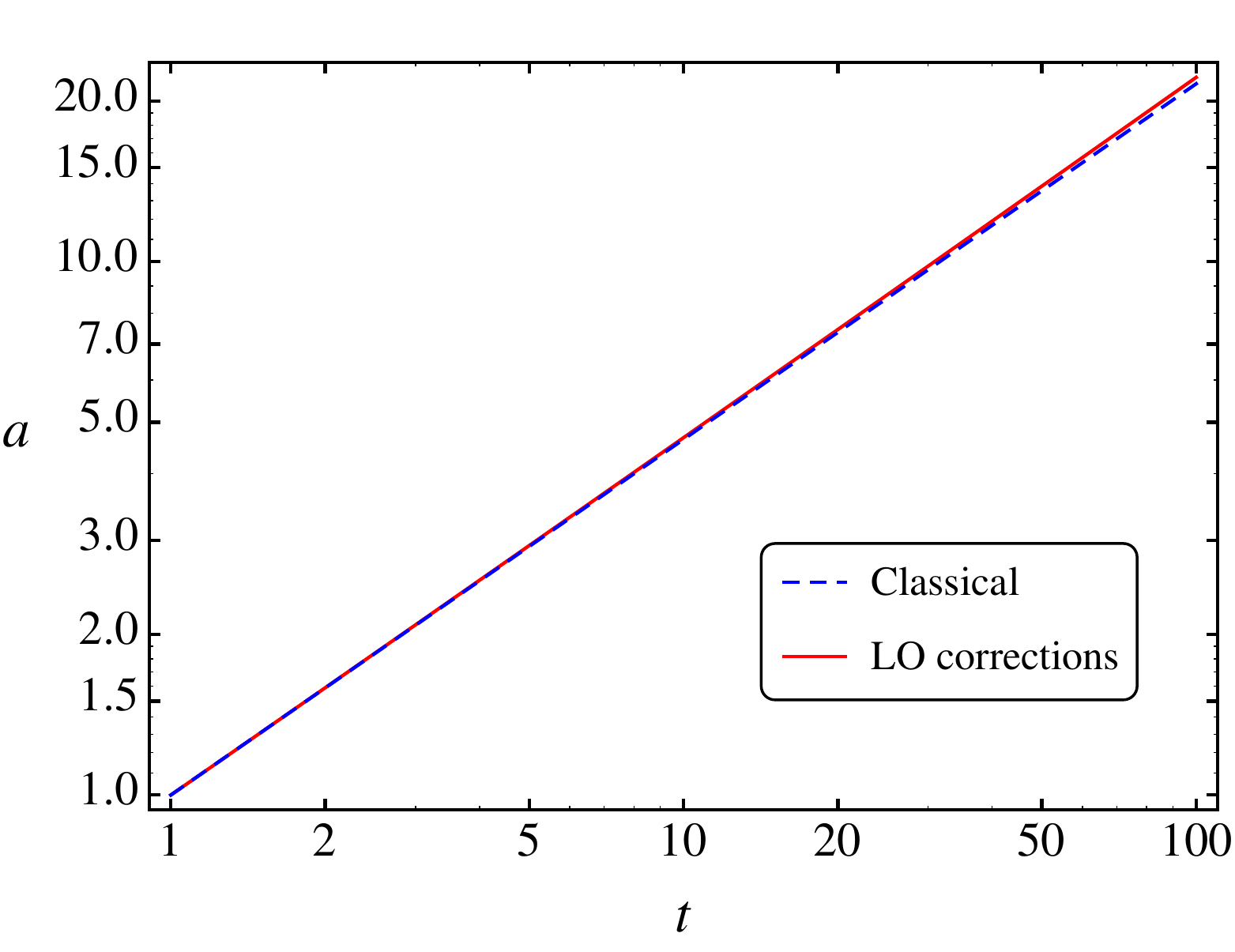}
\hskip 8pt
\includegraphics[width=7.6cm,height=5.5cm]{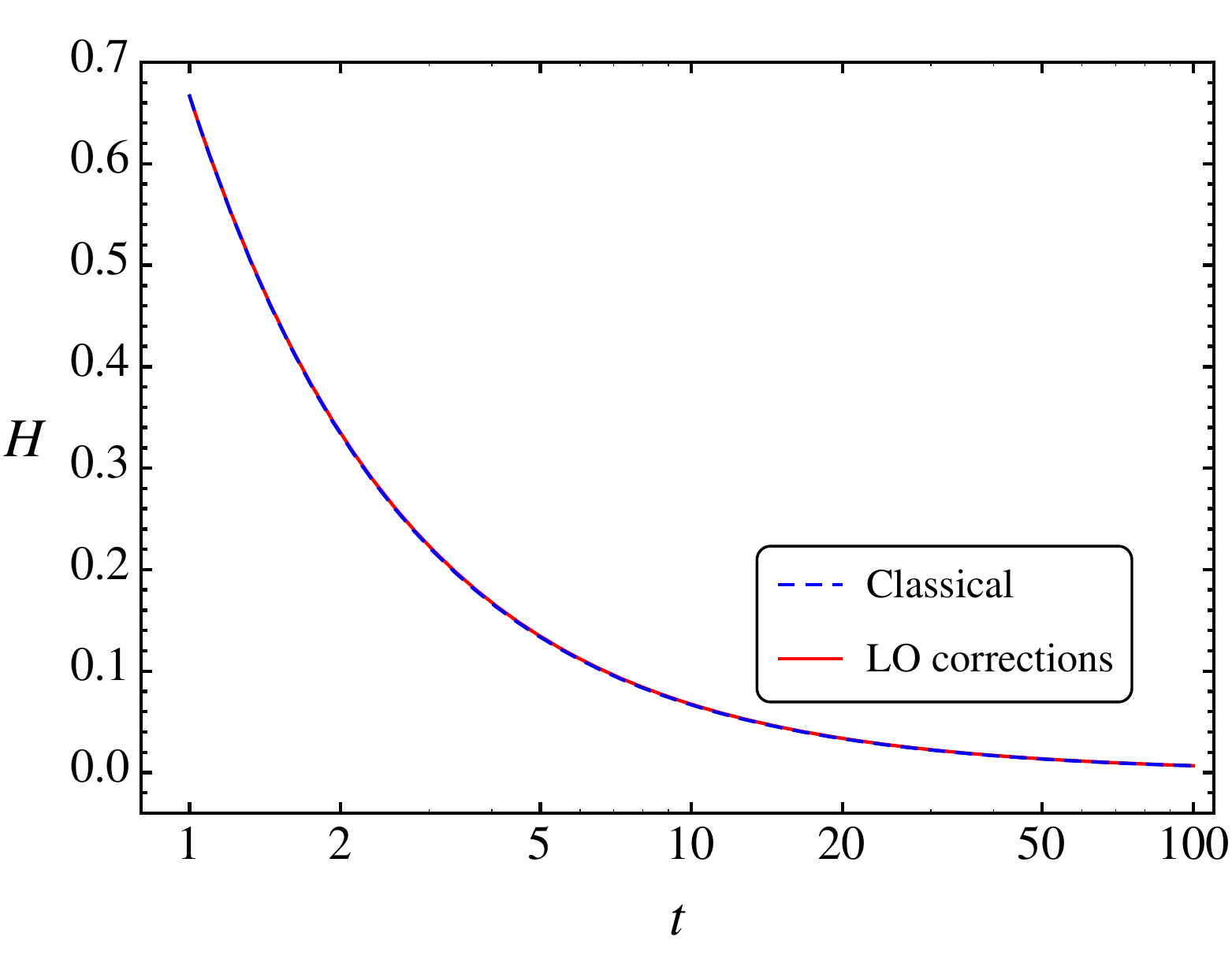}
\caption{Same plot as above but in an expanding matter dominated universe.}
\label{W_a-H-2}
\end{figure}
In Figure \ref{W_a-H-1}, we have plotted the time evolution of the scale factor and the Hubble parameter for the classical solution and the LO corrections for the case of a single minimally coupled scalar ($\chi=0$) and the graviton. It is evident from the figure that the LO corrections for the de Sitter case are significant. For the case of matter as an input solution  we instead find the following effective Friedmann EOM
\begin{equation}
H^2-\frac{128 \pi G}{27} \frac{\beta  m^2}{t^2}\left[1-\left(\frac{t_0}{t}\right)^2-4
 \log \frac{t_0}{t}\right]=\frac{8\pi G}{3}\rho_{m}(t_{0})\left(\frac{t_{0}}{t}\right)^{2}\,,
\end{equation}
which we have solved in the same manner and plotted the solutions for $a$ and $H$ in Figure \ref{W_a-H-2}. As can be seen in this figure, the LO corrections for this case are slightly larger in magnitude than in the previous logarithmic case which can be understood due to the stronger time dependence of the non--local term in this case.  

Finally, we would like to remark that it is possible to integrate the system (\ref{eomw_2_1}) and (\ref{eomw_2_2}) numerically to obtain the evolution of the scale factor and the Hubble parameter without the approximation involved in the iterative solution just exposed. But the outcome clearly shows that the iterative solution is indeed, not only more insightful analytically, but also quantitatively rather precise as we will show in section \ref{comp} when discussing the equation of state parameter.



\subsection{{\small $R\frac{1}{\square^{2}}R$} gravity}\label{maggiore}

This non--local correction is one of the most relevant IR term which has been argued as a simple and consistent non--local extension of GR and thus has been extensively studied in constructing models of dynamical dark energy \cite{Maggiore:2013mea,Maggiore:2014sia}. For this case the non--local action is given by
\begin{equation}
I_{\rm NL}=\int d^{4}x \sqrt{-g}\,R\frac{1}{\square^{2}}R\,.
\label{magg}
\end{equation}
Note that, the effective action $I_{\rm NL}$ is suppressed by a factor of $1/M^2$ compared to the Einstein--Hilbert term, as discussed in eq. (13) of \cite{Codello:2015mba}. From a dimensional point of view, it is easy to see that $\frac{1}{M^2} R\frac{1}{\square^{2}}R$ behaves like a cosmological constant term and therefore, one might expect that it will play an important role in late time cosmology, as we will indeed see later. We also note that scalars and fermions, but not photons, generate this term. In order to obtain the covariant EOM, we start with a variation of (\ref{magg}) which leads to 
\begin{equation}
\delta I_{\rm NL}=\int d^{4}x\left\{ \delta\sqrt{-g}\,R\frac{1}{\square^{2}}R+2\sqrt{-g}\,\delta R\frac{1}{\square^{2}}R-2\sqrt{-g}\,R\frac{1}{\square}\delta\square\left(\frac{1}{\square^{2}}R\right)\right\} \,,
\end{equation}
where we have used $\delta\frac{1}{\square^2}=-\frac{2}{\square}\delta\square\frac{1}{\square^{2}}$ in the last term.
By making use of the basic variations in (\ref{V_1}) and (\ref{var_box}), we obtain the following correction to Einstein's tensor
\begin{eqnarray}
\frac{1}{16\pi G\delta m^{4}}\Delta G_{\mu\nu} & = & \frac{1}{2}g_{\mu\nu}R\left(\frac{1}{\square^{2}}R\right)-2g_{\mu\nu}\left(\frac{1}{\square}R\right)+2\nabla_{\mu}\nabla_{\nu}\left(\frac{1}{\square^{2}}R\right)\nonumber\\
 &  & -\,2R_{\mu\nu}\left(\frac{1}{\square^{2}}R\right)-2\nabla_{\mu}\left(\frac{1}{\square}R\right)\nabla_{\nu}\left(\frac{1}{\square^{2}}R\right)\nonumber\\
 &  & +\,g_{\mu\nu}\nabla^{\alpha}\left(\frac{1}{\square}R\right)\nabla_{\alpha}\left(\frac{1}{\square^{2}}R\right)+g_{\mu\nu}\left(\frac{1}{\square}R\right)^{2}\,.
\end{eqnarray}
To express this equation in a simplified form, as done earlier, we introduce two auxiliary fields, $U=\frac{1}{-\square}R$ and $S=\frac{1}{\square^{2}}R$, and find
\begin{eqnarray}
\frac{1}{16\pi G\delta m^{4}}\Delta G_{\mu\nu} & = & -\,2G_{\mu\nu}S+2g_{\mu\nu}U+2\nabla_{\mu}\nabla_{\nu}S\nonumber\\
 &  & +\,2\nabla_{\mu}U\nabla_{\nu}S-g_{\mu\nu}\nabla^{\alpha}U\nabla_{\alpha}S+\frac{1}{2}g_{\mu\nu}U^{2}\\
-\square U & = & R\\
-\square S & = & U\,.
\end{eqnarray}
One can now immediately write down these equations in the FRW background
\begin{eqnarray}
H^{2}-32\pi G\delta m^{4}\left(2 H^{2}S+H {\dot S}+\frac{1}{2} {\dot H} S-\frac{1}{6} {\dot U}{\dot S}\right) & = & \frac{8\pi G}{3}\rho \label{M_1.1}\\
\ddot U+3H\dot U & = & 6\left(2H^2+\dot H\right) \label{M_1.2}\\
\ddot S+3H\dot S & = & U\,. \label{M_1.3}
\end{eqnarray}
These coupled differential equations form a complete set which can be solved for a given background evolution. It is now easy to rewrite these equations in terms of $'\equiv\frac{d}{d\log a}=\frac{1}{H}\frac{d}{dt}$ to compare them directly with \cite{Maggiore:2014sia}.

%
\begin{figure}[!t]
\includegraphics[width=7.6cm,height=5.5cm]{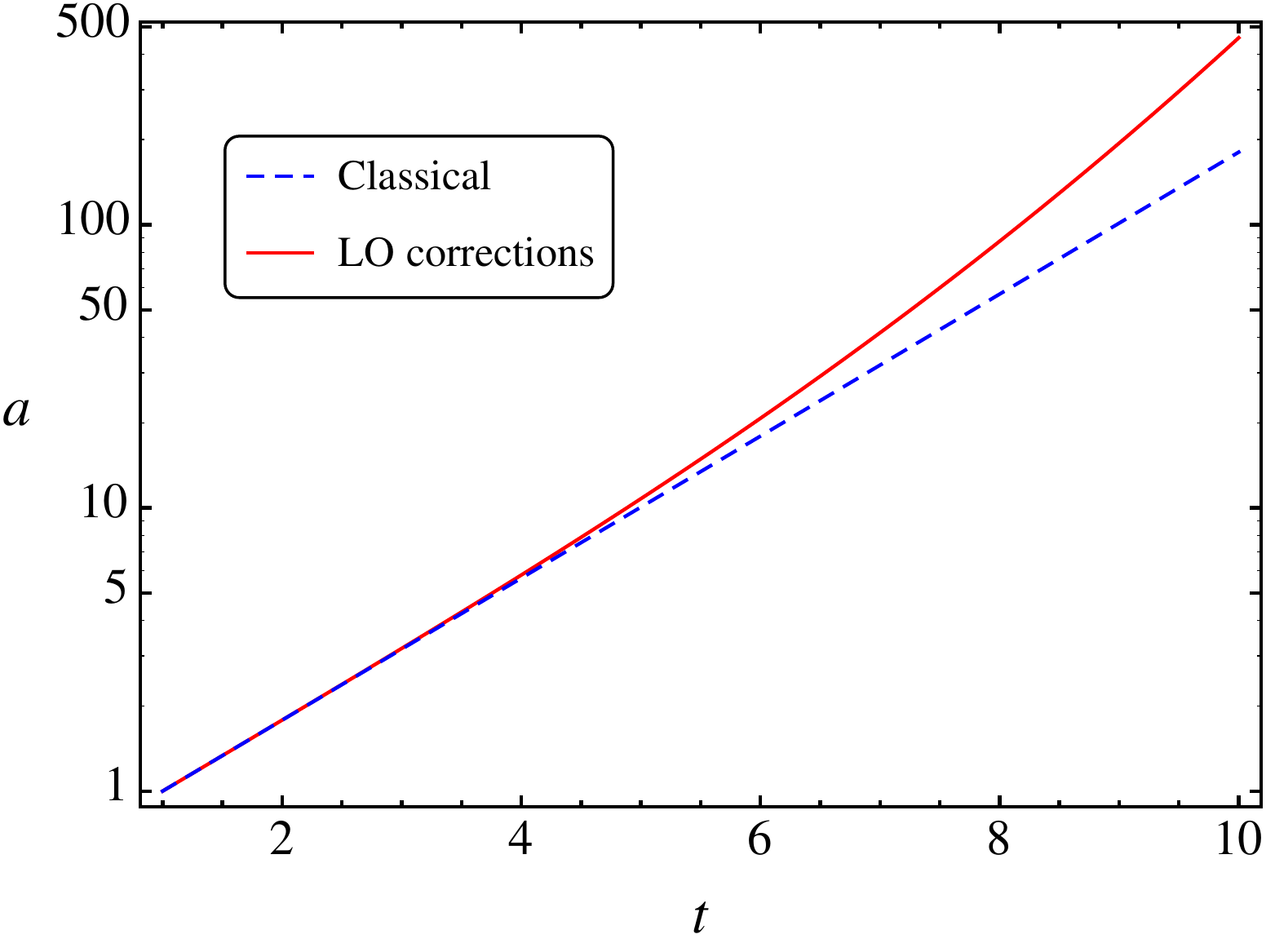}
\hskip 10pt
\includegraphics[width=7.6cm,height=5.5cm]{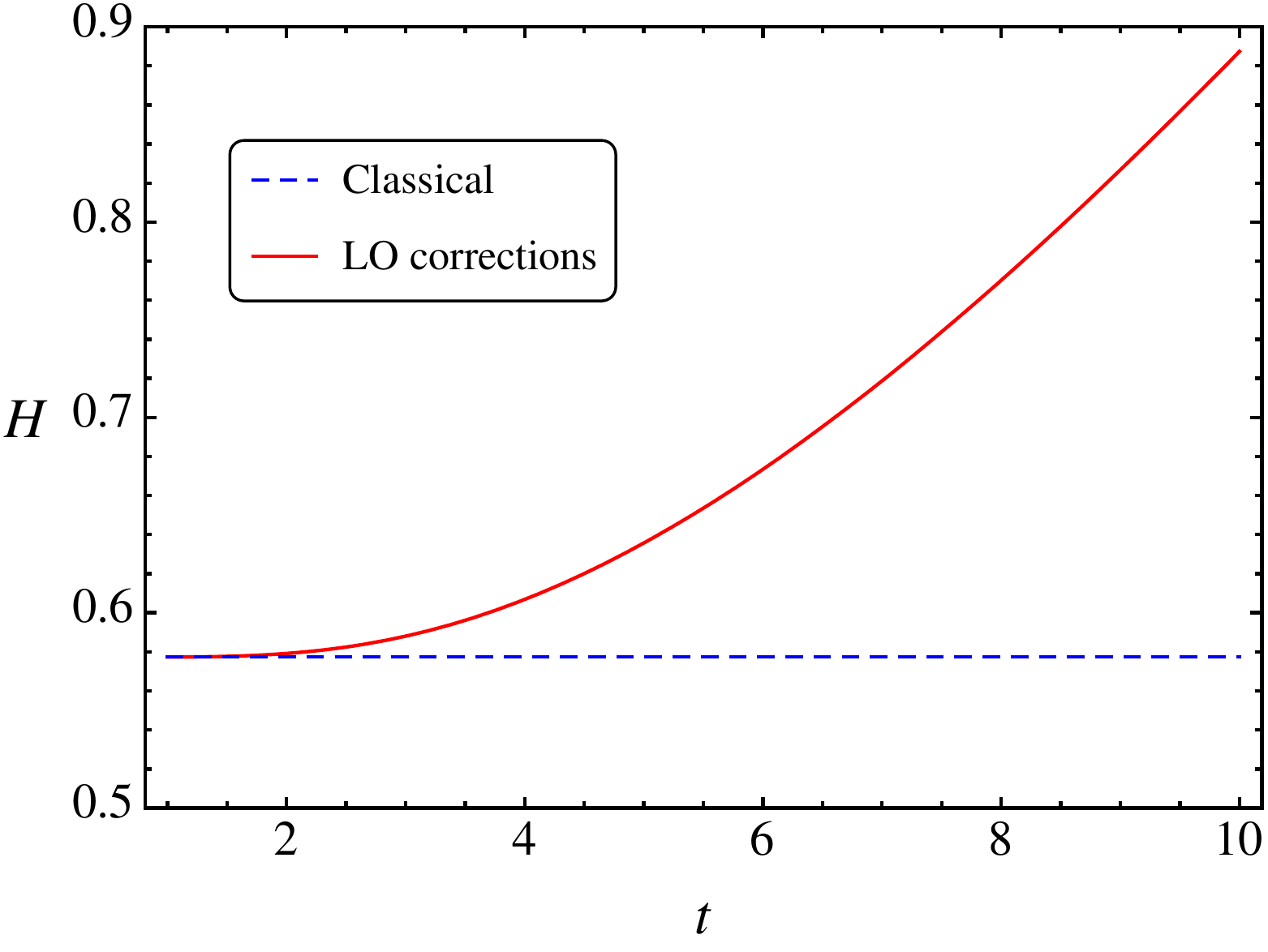}
\caption{The time evolution of the scale factor (left) and the Hubble parameter (right) for the case of $R\frac{1}{\square^2}R$ correction in de Sitter universe with a single minimally coupled scalar ($\chi=0$) and the graviton.}
\label{M_a-H-1}
\vskip 2pt
\includegraphics[width=7.6cm,height=5.85cm]{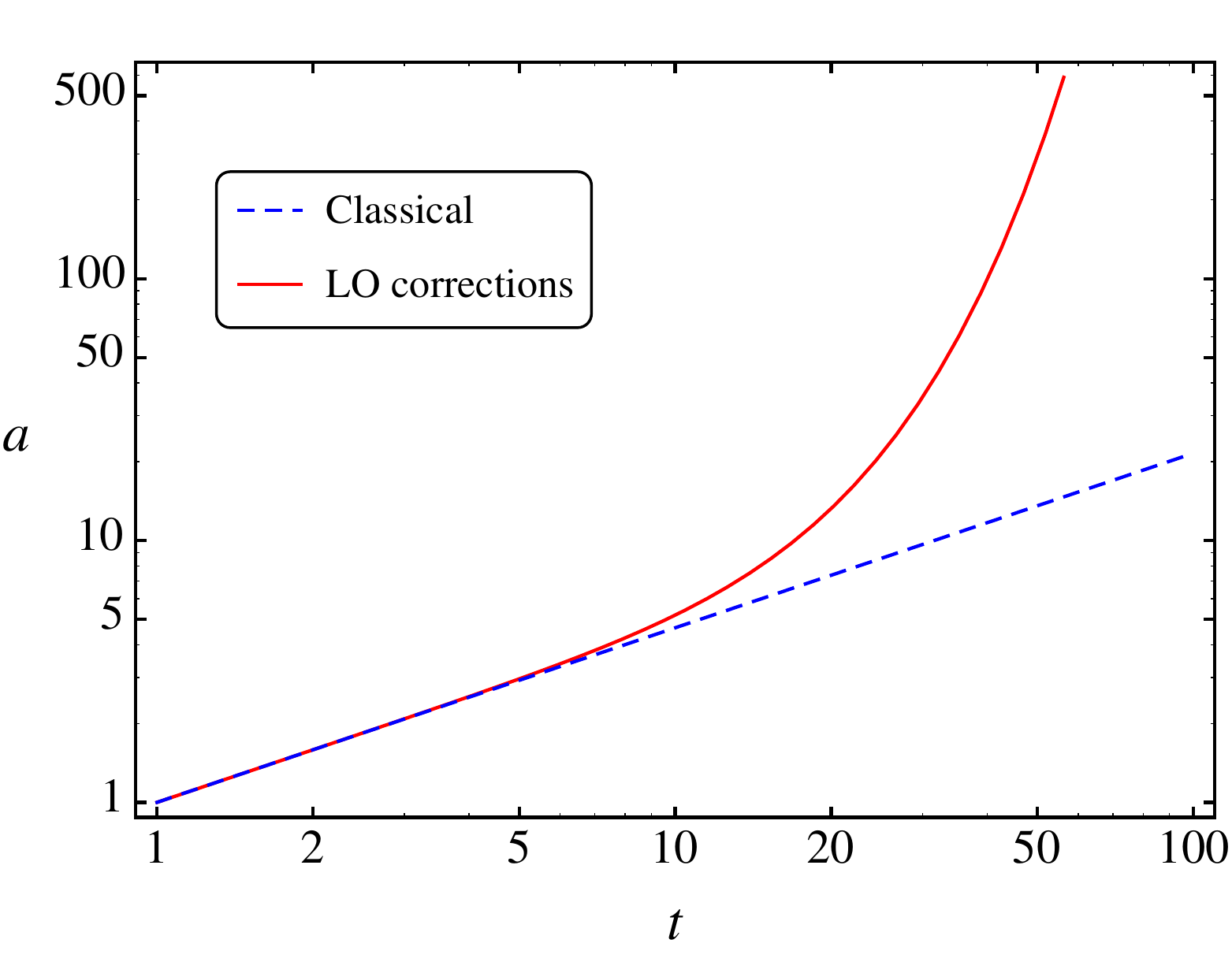}
\hskip 10pt
\includegraphics[width=7.6cm,height=5.5cm]{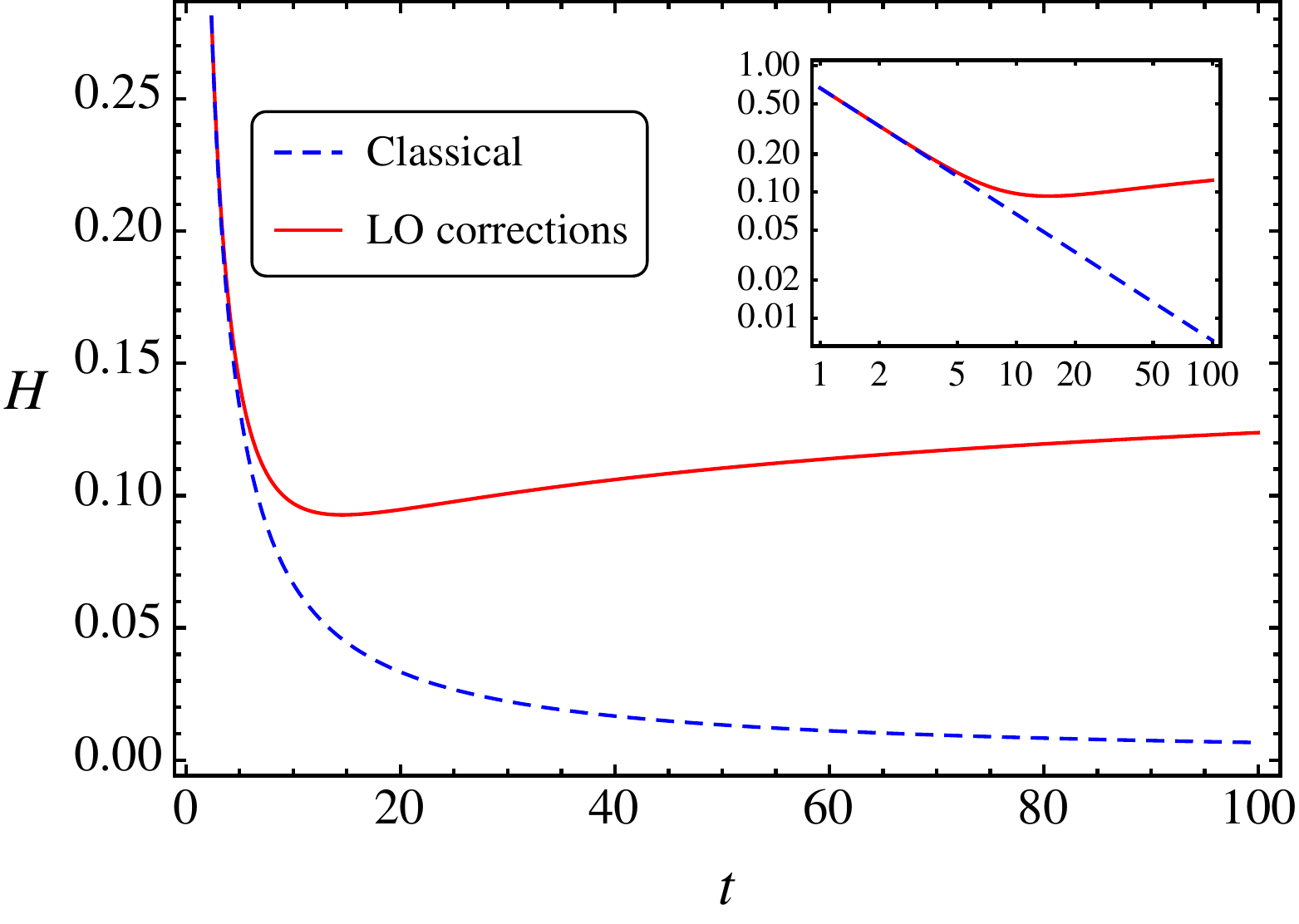}
\caption{Same plot as above but in an expanding matter dominated universe. As shown in the inset of the figure on right, while the Hubble parameter for the classical solution behaves as $\frac{2}{3t}$, the one induced by the LO corrections becomes nearly constant at late times, leading to an accelerated expansion of the universe.}
\label{M_a-H-2}
\end{figure}
As in the earlier case, radiation with $R=0$ is an exact solution for this case while de Sitter and matter are not. We shall solve the closed set of equations (\ref{M_1.1}), (\ref{M_1.2}) and (\ref{M_1.3}) iteratively for a given background expansion of the universe. As mentioned in the previous case, we first solve (\ref{M_1.2}) and (\ref{M_1.3}) for the two scalar potentials $U$ and $S$ with the boundary conditions $U(t_0)={\dot U}(t_0)=0$, $S(t_0)={\dot S}(t_0)=0$ and insert these solutions in (\ref{M_1.1}) to obtain the explicit form of the correction term. On a de Sitter background, the effective Friedmann EOM reads
\begin{eqnarray}
\!\!\!\!H^{2} -\frac{128\pi G}{27}   \delta \,m^4 \bigg\{4-3 \sqrt{3\Lambda }\, (t-t_0)+3 \Lambda  (t-t_0)^2 \bigg.- \left[8- \sqrt{3\Lambda }
  \, (t-t_0)\right]e^{-\sqrt{3\Lambda }\,(t-t_0)} \nonumber\\ 
 \qquad\quad\quad\qquad\qquad\;\bigg.
+\left[4+2\sqrt{3\Lambda }\,(t-t_0)\right]e^{-2 \sqrt{3\Lambda }\,(t-t_0)} \bigg\} = \frac{\Lambda}{3}\,.\label{M_1}
\end{eqnarray}
The correction term in this case contains a constant term, a linear and quadratic term as well as exponentially decaying terms of characteristic timescale $1/\sqrt{3\Lambda}$.  As earlier, the correction term vanishes as $t \to t_0$ as well as $\Lambda \to 0$, as expected. At late times, the exponentially decaying terms can be ignored compared to the others and thus, the most relevant term in the deep IR regime will be the quadratic one.  
As said, this non--local term is not generated by photons but only by scalars and fermions and therefore, the expression for $\delta$ is given by
\begin{equation}
\delta=\frac{1}{(4 \pi)^2} \left[-\frac{15-6\chi- \chi^2}{144}N_0+\frac{1}{6}N_{1/2}+\frac{17}{24}\right]\,.\label{M_2}
\end{equation}

\!\!We now solve this equation numerically and plot the time evolution of the scale factor and the Hubble parameter for the classical solution and the LO corrections for the case of a single minimally coupled scalar ($\chi=0$) and the graviton in Figure \ref{M_a-H-1}. We find that due to the stronger time dependence of the correction term, the LO corrections are significantly large at late times. To understand the effect of the correction term in the case of matter, we first write down the effective Friedmann EOM
\begin{eqnarray}
H^{2}-\frac{32\pi G}{81} \delta \,m^4 \Bigg\{\!-\frac{119}{3}-26 \log \frac{t_0}{t} +\frac{t_0}{t}  \left(\frac{130}{3}-8 \log \frac{t_0}{t}\right)\Bigg. \nonumber
\\  \qquad\qquad\qquad\, \Bigg.
-3\left(\frac{t_0}{t}\right)^2+\frac{2}{3}\left(\frac{t_0}{t}\right)^3-\frac{4}{3} \left(\frac{t_0}{t}\right)^4\Bigg\} 
= \frac{8\pi G}{3}\rho_{m}(t_{0})\left(\frac{t_{0}}{t}\right)^{2},
\end{eqnarray}
and then solve it numerically. Note that, the correction term correctly vanishes as $t \to t_0$. In Figure \ref{M_a-H-2}, we have plotted the numerical solution for the scale factor and the Hubble parameter for the case of a single minimally coupled scalar ($\chi=0$) and the graviton. We find that the LO corrections are quite significant for this case as well.  More interestingly, we notice that, at late times, the Hubble parameter induced by the LO corrections becomes nearly constant  and leads to an accelerated expansion of the universe, as shown clearly in the inset of the right Figure \ref{M_a-H-2}.

We shall further discuss this interesting behavior in the next section when discussing the equation of state parameter.
We will also show that the iterative solution just exposed is, as in the previous case, quite a good approximation in comparison to the full numerical solution in both the cases. 

Before we close this section, an interesting remark about the corrections in de Sitter for $\frac{1}{-\square}$ and $\frac{1}{\square^2}$ terms is in order. It can be seen from (\ref{W_1}) and (\ref{M_1}) that the corrections due to the non--local terms in de Sitter lead to an effective time dependent cosmological constant in a consistent way which is a distinct outcome of our formalism and it is induced in a completely covariant set up without breaking the diffeomorphism invariance\footnote{Note that the canonical method of calculating one-loop quantum corrections also does not break the diffeomorphism symmetry. However, the canonical method is not manifestly diffeomorphism invariant while our covariant approach is.}. This leads to interesting phenomenological consequences, as discussed in \cite{Peebles:1987ek,Freese:1986dd,Carvalho:1991ut,Overduin:1998zv}.



\section{A minimal comparison of non--local corrections}\label{comp}

In this section we quickly compare the iterative solutions obtained in section \ref{nlocalg} for the non--local corrections predicted by the EFT of gravity in order to understand their relative role in the evolution of the universe at late times. We can treat these correction terms as an effective dark energy component and write the general Friedmann EOM in the following simple form
\begin{equation}
H^{2}=\frac{8\pi G}{3} (\rho+\rho_{\rm de})\,.
\label{gefe}
\end{equation}
\begin{figure}[!t]
\begin{center}
\includegraphics[width=14.0cm,height=8.0cm]{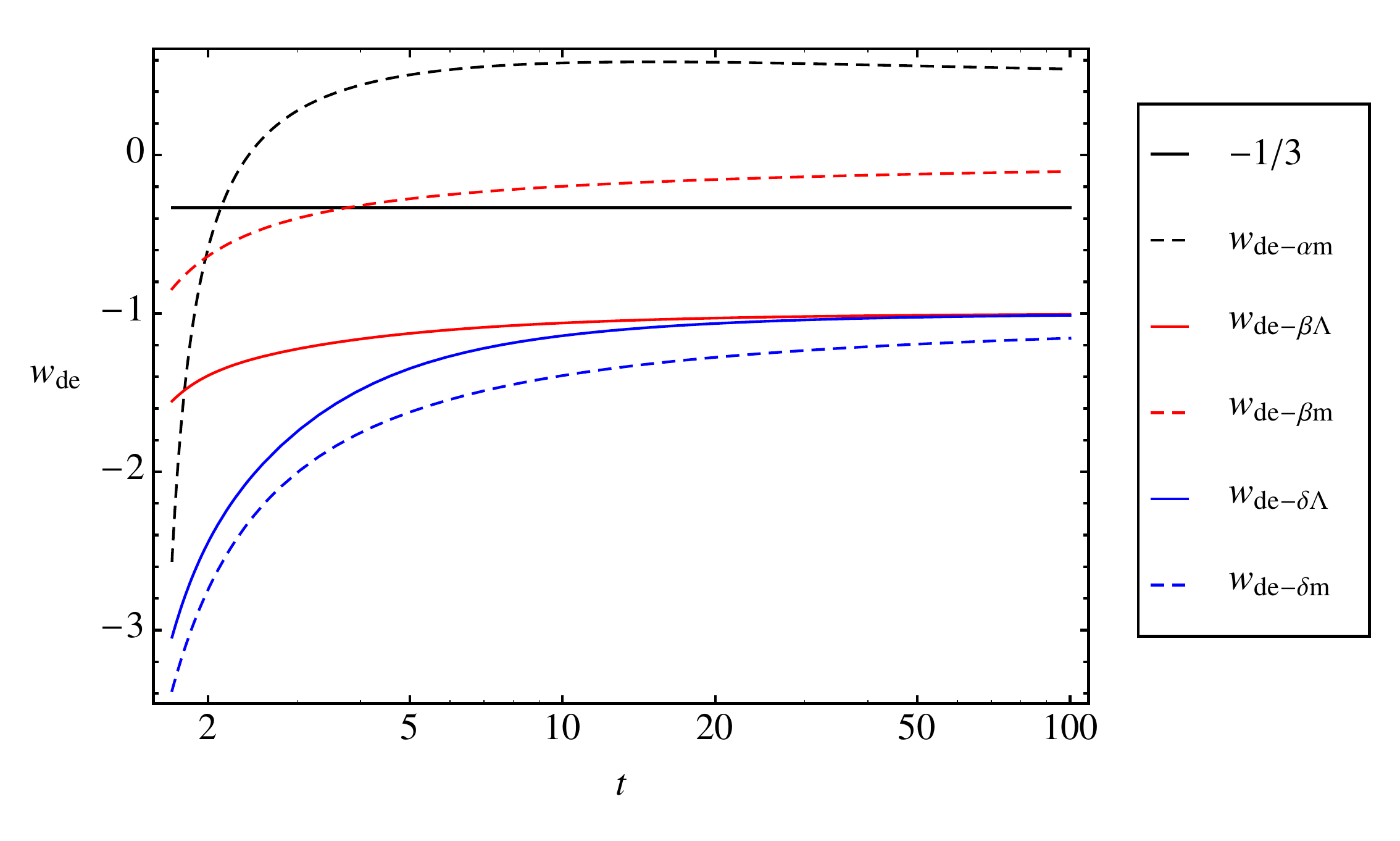}
\caption{The equation of state parameter $w_{\rm de}$ is plotted as a function of time corresponding to the different non--local correction terms. It is evident from the figure that the corrections induced by matter for the case of the $R\frac{1}{\square^{2}}R$ term can indeed lead to an accelerated expansion of the universe with $w_{\rm de}<-1$. Note that the behavior of $w_{\rm de}$ for different cases is independent of the respective coefficients $\alpha$, $\beta$ and $\delta$. }
\label{compare}
\end{center}
\end{figure}
The effective dark energy densitiy $\rho_{\rm de}$ for each of  the various LO non--local corrections can be read off from the previous section and have the following form
\begin{eqnarray}
\rho_{\rm de}^{\alpha,m}(t) &=&  \frac{32\alpha}{\,t^{4}}\left[\log m t+\log\left(\frac{t}{t_{0}}-1\right)+\frac{2}{3}\left(\frac{t}{t_{0}}-1\right)\right] \\
\nonumber\\
\rho_{\rm de}^{\beta,\Lambda}(t) &=&  \frac{16}{3}  \beta m^2 \Lambda \left(1 +\sqrt{3 \Lambda}\,(t-t_0)-  e^{-2 \sqrt{3 \Lambda }\, (t-t_0)}\right) \\
\nonumber\\
\rho_{\rm de}^{\beta,m}(t) &=&  \frac{16}{9} \beta  m^2 \frac{1}{t^2}\left[1-\left(\frac{t_0}{t}\right)^2-4
 \log \frac{t_0}{t}\right] \\
\nonumber\\
\rho_{\rm de}^{\delta, \Lambda}(t) &=&  \frac{16}{9}\delta m^4 \bigg\{4-3 \sqrt{3\Lambda }\, (t-t_0)+3 \Lambda  (t-t_0)^2 \bigg.\nonumber\\
&& \bigg.- \left[8- \sqrt{3\Lambda }\,(t-t_0)\right]e^{-\sqrt{3\Lambda }\,(t-t_0)} +\left[4+2\sqrt{3\Lambda }\,(t-t_0)\right]e^{-2 \sqrt{3\Lambda }\,(t-t_0)} \bigg\}\\
\nonumber\\
\rho_{\rm de}^{\delta, m}(t) &=& \frac{4}{27} \delta m^4 \Bigg\{\!-\frac{119}{3}-26 \log \frac{t_0}{t} +\frac{t_0}{t}  \left(\frac{130}{3}-8 \log \frac{t_0}{t}\right)\Bigg. \nonumber \\
&&\Bigg. -3\left(\frac{t_0}{t}\right)^2+\frac{2}{3}\left(\frac{t_0}{t}\right)^3-\frac{4}{3} \left(\frac{t_0}{t}\right)^4\Bigg\}\,.
\label{rhode}
\end{eqnarray}
Here, the superscripts on each term indicate the corresponding correction term and the input classical solution. For instance, $\rho_{\rm de}^{\alpha,m}$ indicates the correction term for the logarithmic case with matter solution as an input.
Note that, as also discussed earlier, the correction term for the logarithmic case vanishes for the de Sitter expansion (\ref{dsa}) as all the radiation cases. The aim is to understand the late time evolution of these corrections and see if they can mimic an effective dark energy component, leading to an accelerated expansion of the universe. The fact that the correction terms are covariantly conserved  leads to the continuity equation for $\rho_{\rm de}$,
\begin{equation}
\dot \rho_{\rm de}+ 3H (1+w_{\rm de}) \rho_{\rm de}=0\,,
\end{equation}
where $w_{\rm de}$ is the effective equation of state parameter for the dark energy component. This equation immediately leads to
\begin{equation}
w_{\rm de}=-1-\frac{1}{3H}\left(\frac{\dot \rho_{\rm de}}{\rho_{\rm de}}\right)\,.
\end{equation}

\!It is interesting to note that while $\rho_{\rm de}$ for different cases depends on the respective coefficients $\alpha$, $\beta$ and $\delta$, the equation of state parameter $w_{\rm de}$ is indeed independent of them. One should typically plot the evolution of the total equation of state parameter $w$ for a system with multiple fields or fluids but it turns out that, using the continuity equations for the total energy density, matter and the induced dark energy component, one can write down an expression for $w$ as
\begin{equation}
w=\frac{\rho_{\rm de}}{\rho_{\rm m}+\rho_{\rm de}}w_{\rm de}\,,
\end{equation}
which tells us that $w \approx w_{\rm de}$ for $\rho_{\rm m} \ll \rho_{\rm de}$ at late times and therefore, we plot $w_{\rm de}$ as a function of time for the different cases, as shown in Figure \ref{compare}.  While the corrections induced by matter in the cases of logarithmic and $\frac{1}{\square}$ terms lead to $w_{\rm de}>0$ and $-1/3<w_{\rm de}<0$, respectively, it is a naive expectation that the correction terms induced by a de Sitter solution in any case will lead to $w_{\rm de}\leq -1$, which is what we have found for the cases of $\frac{1}{\square}$ and $\frac{1}{\square^2}$ terms. Moreover, the corrections induced by matter in the case of $\frac{1}{\square^{2}}$ are indeed the most interesting ones as they generically lead to a phantom like behavior with $w_{\rm de}<-1$ and can indeed explain the current acceleration of the universe without the need for a quintessence field \cite{Maggiore:2013mea,Maggiore:2014sia}. 
\begin{figure}[!t]
\includegraphics[width=7.6cm,height=5.5cm]{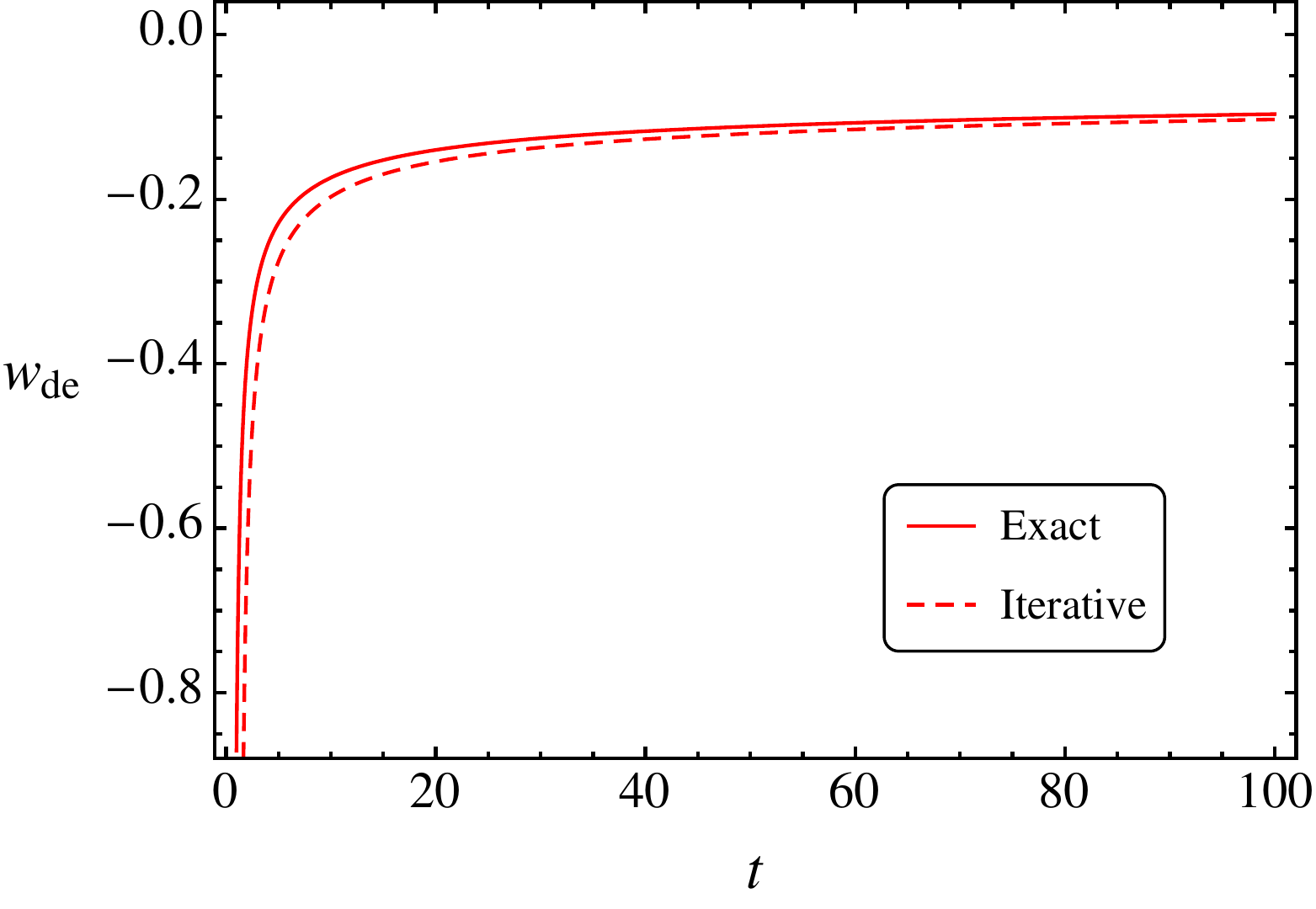}
\hskip 6pt
\includegraphics[width=7.6cm,height=5.5cm]{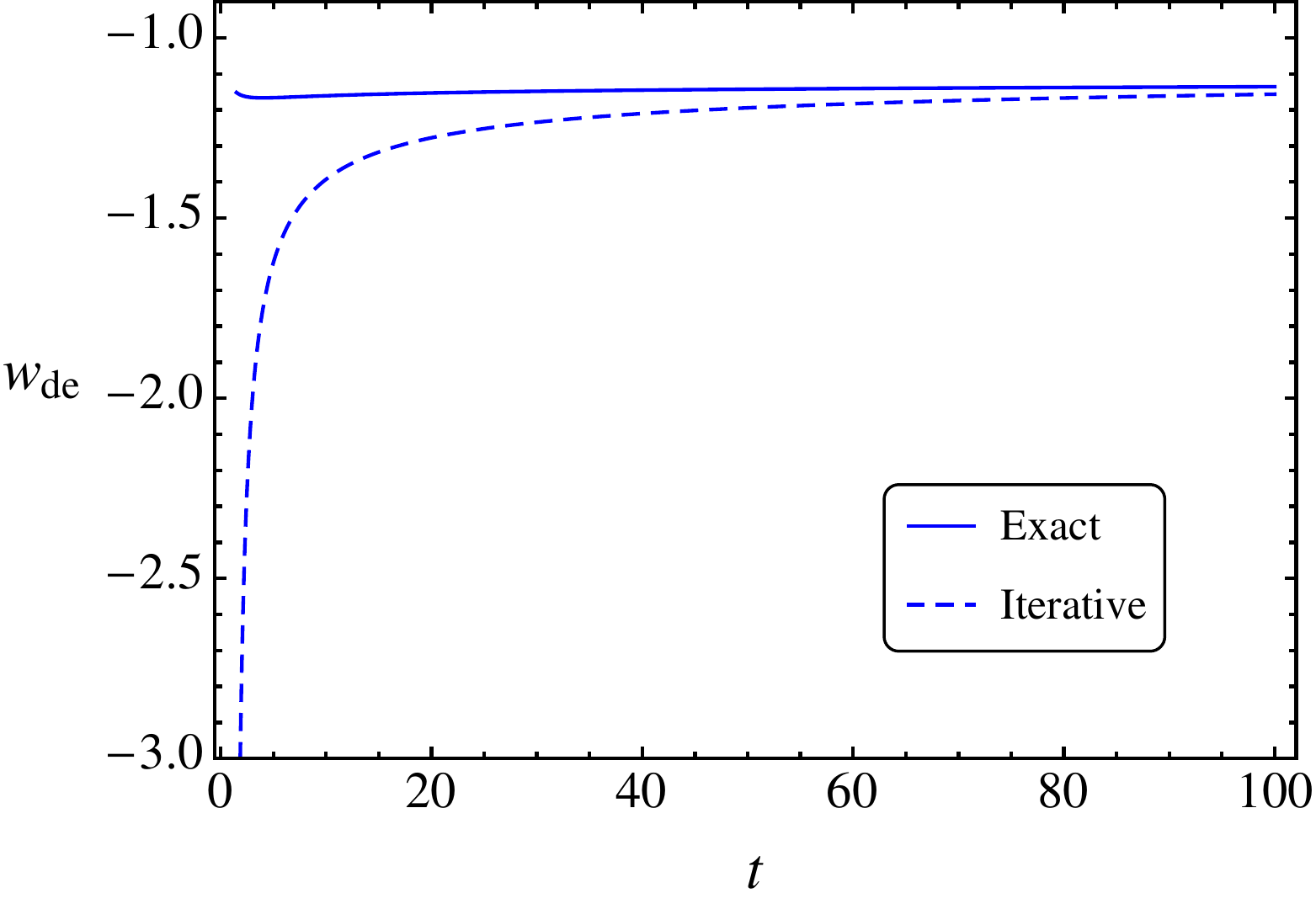}
\caption{We compare the exact and iterative solutions for $w_{\rm de}$ for the $R\frac{1}{-\square}R$ (left) and $R\frac{1}{\square^2}R$ (right) correction terms.  We found that the iterative solution is a very good approximation to the exact solution in both the cases asymptotically.}
\label{compare-sols}
\end{figure}

Before we close this section, a few comments are in order about the comparison between the exact and iterative solutions for the cases of $R\frac{1}{-\square}R$ and $R\frac{1}{\square^2}R$ corrections. We have plotted both the solutions for these two cases in Figure \ref{compare-sols}. It is clear from the figure that the iterative solution is indeed a good approximation to the exact solution in both these cases and seem to capture the subdominant effects quite well. Therefore, we can argue that the iterative solution serves as a good approximation to the exact solution in all the cases. Hence, it is a very useful tool to obtain the solutions  in cases wherein the system of equations can not be solved exactly.


\section{Discussion and Outlook}\label{discuss}

In our previous paper \cite{Codello:2015mba}, we had constructed an EFT of gravity in a complete covariant manner by means of an expansion in the inverse powers of Planck mass and identified the classical theory together with the leading and next--to--leading quantum corrections. We then computed the leading order corrections to the quadratic order in curvature and found that these corrections in the effective action can be classified into two categories: local and non--local. In the FRW spacetime, the leading local correction is given by the $R^2$ term and its coupling is a free parameter which must be fixed by observations.  The non--local sector instead contains different terms relevant at low energy and their coefficients are completely determined by the EFT of gravity once the matter content of the theory is specified. Since the $R^2$ term is a UV term and is primarily relevant during the early evolution of the universe, an inflationary scenario based on it can be constructed \cite{Starobinsky:1980te}. As we discussed, this scenario can be naturally accommodated in framework of the EFT of gravity. 

In this paper, we have focussed our attention on studying the cosmological consequences of the non--local terms when each of them is combined with the Einstein--Hilbert action. 
We have explained how to derive the covariant EOM for all the cases and also how to correctly project these expressions on the FRW background, taking care in explaining how to represent the different non--local contributions. It turns out that these terms are not so difficult to handle and can be easily dealt with in our covariant formalism. 
We have then studied both the analytical and iterative solutions in all the cases and also commented on the comparison between the iterative and numerical solutions whenever possible. 
We have compared the evolution of the scale factor and the Hubble parameter induced by the non--local corrections to the classical solutions in all the cases and found that the imprints due to the $R\frac{1}{\square^2}R$ term have the strongest effect among all the terms considered. Finally, we compared all these contributions after rewriting them in terms of an effective dark energy component characterized by an equation of state parameter and found that $R\frac{1}{\square^2}R$ term can indeed drive an accelerated expansion of the universe at the present epoch \cite{Maggiore:2013mea,Maggiore:2014sia}. 

One of the key result of this paper is that all the non--local terms in the EFT of gravity arise in a consistent way from the first principles and have precise coefficients which depend ultimately on the matter content. 
Over the last few years, a large class of phenomenological models containing one or more of these non--local terms have been proposed in the literature.  Here we have shown that all such terms appear with precise coefficients which hints at the possibility to falsify these phenomenological scenarios. Also, all these non--local terms generally appear together in the EFT of gravity unless certain matter content of the theory renders one of them non--vanishing while killing the others. But, in any case, the logarithmic term is {\it always} present and therefore, none of the other non--local terms can be considered alone.

One of the opportunities offered by the covariant EFT of gravity is the inclusion of the cosmological constant in a consistent way. The combined effect of the   cosmological constant and the non--local terms is to effectively induce a time--dependence of the former \cite{Peebles:1987ek,Freese:1986dd,Carvalho:1991ut,Overduin:1998zv}. We emphasize that this is an interesting outcome in our covariant formalism which is manifestly diffeomorphism invariant contrary to the canonical method. We plan to come back to it in the future. 

An interesting possibility would be to study a scenario wherein the leading local correction term is combined with the IR relevant non--local terms. Such a scenario would lead to an early inflationary epoch together with an accelerating phase at the present epoch thereby unifying inflation and dark energy in a single framework by means of pure gravitational corrections without the usual need of a scalar field \cite{Codello:2016neo,Codello:2016elq,Codello:2016xhm}. Furthermore, the next natural step would be to study the evolution of cosmological perturbations in the EFT of gravity. As mentioned earlier, the Weyl contributions to the effective action will play a pivotal role in this context and it will be interesting to look for such imprints in cosmological observables.


\paragraph*{Acknowledgments}

A.C. would like to thank John F. Donoghue, Basem Kamal El--Menoufi and Roberto Percacci for useful discussions.
R.K.J. acknowledges financial support from the Danish Council for Independent Research in Natural Sciences for part of the work.
The CP$^3$-Origins centre is partially funded by the Danish National Research Foundation, grant number DNRF90.


\appendix

\section{Non--local operators on FRW}\label{appA}

We are interested in providing a precise meaning to the Green's function  $\frac{1}{-\square+m^2}$ and the operator $\log\frac{-\square}{\mu^{2}}$.

\subsection{Green's function}

To compute the Green's function on the FRW background, we need to solve
\begin{equation}
(-\square+m^2) G^{\rm FRW}(t-t';m)=\tilde \delta(t-t')\,,\label{G_1}
\end{equation}
where
\begin{equation}
\square=-\frac{1}{a^{3}(t)}\frac{d}{dt}\left(a^{3}(t)\frac{d}{dt}\right)\,,\label{G_2}
\end{equation}
and
\begin{equation}
\tilde{\delta}(t-t')\equiv\frac{\delta(t-t')}{\sqrt{a^{3}(t)a^{3}(t')}}\,,\label{G_3}
\end{equation}
is the covariant delta function, normalized such that $ \int dt'a^{3}(t')\tilde{\delta}(t-t')=1\,.$
Plugging (\ref{G_2}) and (\ref{G_3}) in (\ref{G_1}) leads to 
\begin{equation}
\frac{1}{a^{3}(t)}\frac{d}{dt}\left(a^{3}(t)\frac{d}{dt}G^{\rm FRW}(t-t';m)\right)+m^{2}G^{\rm FRW}(t-t';m)=\frac{\delta(t-t')}{\sqrt{a^{3}(t)a^{3}(t')}}\,.\label{G_20}
\end{equation}
The full kernel in a general FRW spacetime is not realistically tractable. So as an approximation we drop the time derivatives of the scale factor and rewrite (\ref{G_20})  as
\begin{equation}
\left(\frac{d^{2}}{dt^{2}}+m^{2}\right)\left[a^{3}(t)G^{\rm FRW}(t-t';m)\right]+O(\dot{a})=\delta(t-t')\,.\label{G_21}
\end{equation}
Thus, for a slowly varying scale factor, we can use the {\it retarded} (or casual) flat space massive Green's function
\begin{equation}
G^{\rm flat}_{-}(t;m)=\theta(t)\frac{\sin mt}{m}\,,\label{G_12.1}
\end{equation}
to write the following symmetric form for the solution in (\ref{G_21})
\begin{equation}
G^{\rm FRW}_{-}(t-t';m)=\frac{\theta(t-t')}{\sqrt{a^{3}(t)a^{3}(t')}}\frac{\sin m(t-t')}{m}+O(\dot{a})\,.\label{G_22}
\end{equation}
In the massless limit we find
\begin{equation}
G^{\rm FRW}_{-}(t-t';0)=\frac{\theta(t-t')(t-t')}{\sqrt{a^{3}(t)a^{3}(t')}}+O(\dot{a})\,.\label{G_23}
\end{equation}
The general FRW Green's function can be found using the Strum--Liouville theory.


\subsection{The $\log\frac{-\square}{\mu^{2}}$ distribution}

The distribution $L^{\rm FRW}\equiv\log\frac{-\square}{\mu^{2}}$ is precisely defined as
\begin{equation}
L^{\rm FRW}(t-t')=\int_{0}^{\infty}ds\left[\frac{1}{\mu^{2}+s}-G^{\rm FRW}(t-t';\sqrt{s})\right]\,,\label{L_1}
\end{equation}
where different boundary conditions on the Green's function $G^{\rm FRW}$ define different kernels $L^{\rm FRW}$.
We are interested in the retarded kernel $L^{\rm FRW}_{-}$ obtained by plugging (\ref{G_22}) in (\ref{L_1}). We need to evaluate the following integral of the retarded flat space Green's function
\begin{eqnarray*}
I(t)  \equiv  \int_{0}^{\infty}ds\, G^{\rm flat}_{-}(t;\sqrt{s})
  =  \theta(t)\int_{0}^{\infty}ds\frac{\sin\sqrt{s}\,t}{\sqrt{s}}
  =  2\theta(t)\int_{0}^{\infty}dy\,\sin y\, t\,,
\end{eqnarray*}
where we have changed variables to $y=\sqrt{s}$, $dy=\frac{ds}{2\sqrt{s}}$.
The last integral is meaningful only if interpreted as a distribution and in particular we have\footnote{The calculation steps are as follows
\begin{equation*}
\int_{0}^{\infty}dy\,\sin y\, t  =  \lim_{\epsilon\rightarrow0}\int_{0}^{\infty}dy\, e^{-\epsilon y}\sin y\, t\\
 =  \frac{1}{2}\lim_{\epsilon\rightarrow0}\left[\frac{1}{t+i\epsilon}-\frac{1}{t-i\epsilon}\right]\\
 = \mathcal{P}\frac{1}{t}\,,
\end{equation*}
where in the last step we used $ \lim_{\epsilon\rightarrow0}\frac{1}{t\pm i\epsilon}=\mathcal{P}\frac{1}{t}\mp i\pi\delta(t)$.}
\begin{equation}
\int_{0}^{\infty}dy\,\sin y\, t=\mathcal{P}\frac{1}{t}\,,\label{L_2}
\end{equation}
where the principal value distribution is defined as
\begin{equation}
\left(\mathcal{P}\frac{1}{f}\right)[g]\equiv\lim_{\epsilon\rightarrow0}\sum_{x_{i}|f(x_{i})=0}\left[\int_{-\infty}^{x_{i}-\epsilon}dx\frac{g(x)}{f(x)}+\int_{x_{i}+\epsilon}^{\infty}dx\frac{g(x)}{f(x)}\right]\,.\label{G_11}
\end{equation}
Thus, bringing the step function inside the principal value, we get
\begin{equation}
I(t-t')=2\mathcal{P}\frac{\theta(t-t')}{t-t'}\,.\label{L_3}
\end{equation}
Integrating against a test function gives
\begin{eqnarray*}
\int_{-\infty}^{\infty}dt'\, I(t-t')f(t')
 & = & 2\mathcal{P}\int_{-\infty}^{\infty}dt'\frac{\theta(t-t')}{t-t'}f(t')\\
 & = & 2\lim_{\epsilon\rightarrow0}\left[\int_{-\infty}^{t-\epsilon}dt'\frac{\theta(t-t')}{t-t'}f(t')+\underbrace{\int_{t+\epsilon}^{\infty}dt'\frac{\theta(t-t')}{t-t'}f(t')}_{=0}\right]\\
 & = & 2\lim_{\epsilon\rightarrow0}\int_{-\infty}^{\infty}dt'\frac{\theta(t-t'-\epsilon)}{t-t'}f(t')\,,
\end{eqnarray*}
or simply
\begin{equation}
\mathcal{P}\frac{\theta(t-t')}{t-t'}=\lim_{\epsilon\rightarrow0}\frac{\theta(t-t'-\epsilon)}{t-t'}\,.\label{L_4}
\end{equation}
The other integral in (\ref{L_1}) needs a regularization
\begin{equation}
\int_{0}^{\infty}ds\frac{1}{\mu^{2}+s}=\lim_{\epsilon\rightarrow0}\int_{0}^{1/\epsilon^{2}}ds\frac{1}{\mu^{2}+s}=-2\lim_{\epsilon\rightarrow0}\log\mu\,\epsilon\,.
\end{equation}
As a distribution this is represented as  $-2\delta(t-t')\log\mu\epsilon$ and combining with (\ref{L_4}) finally gives the flat space kernel
\begin{equation}
L^{\rm flat}_{-}(t-t')=-2\lim_{\epsilon\rightarrow0}\left[\frac{\theta(t-t'-\epsilon)}{t-t'}+\delta(t-t')\log\mu\,\epsilon\right]\,.\label{L_5}
\end{equation}
Using (\ref{G_22}) we can write the following expression for the FRW retarded kernel
\begin{equation}
L_{-}^{\rm FRW}(t-t')=-2\lim_{\epsilon\rightarrow0}\left[\frac{\theta(t-t'-\epsilon)}{\sqrt{a^{3}(t)a^{3}(t')}}\frac{1}{t-t'}+\frac{\delta(t-t')}{\sqrt{a^{3}(t)a^{3}(t')}}\log\mu\,\epsilon\right]\,,\label{L_7}
\end{equation}
which is valid for a slowly varying scale factor.


\bibliographystyle{JHEP}
\bibliography{EFT_GR_II_Bibliography}


\end{document}